\documentclass[useAMS,usenatbib]{mn2e}

\usepackage {graphicx}
\usepackage {times}

\def\apj{\,{\rm ApJ}}

\title[]{Optical Properties of 4248 IRAS Galaxies}
\author[T. Goto]
{Tomotsugu Goto$^{1}$\thanks{E-mail:tomo@jhu.edu}
  \\
  $^{1}$ Department of Physics and Astronomy, The Johns Hopkins
  University, 3400 North Charles Street, Baltimore, MD 21218-2686, USA
}

\begin{document}
\maketitle


%





\begin{abstract}\label{abstract}
We have investigated optical properties of 4248 infrared galaxies (IRGs) by positionally matching data from the Infrared Astronomical Satellite (IRAS) and the Sloan Digital Sky Survey Data Release 3 (SDSS DR3). Due to the large sky area coverage of both surveys, we have obtained an opportunity to study an unprecedentedly large number of 4248 infrared galaxies spanning four orders of infrared luminosity range $10^{9}L_{\odot} \leq L_{8-1000\mu m} \leq 10^{13.57}L_{\odot}$. Our sample includes three hyper luminous infrared galaxies ($L_{8-1000\mu m} >10^{13}L_{\odot}$) and the large number of 178 ultra luminous infrared galaxies ($10^{12}L_{\odot} < L_{8-1000\mu m} \leq 10^{13}L_{\odot}$). 
 In addition, it is important to have statistical number of lower luminosity infrared galaxies ($L_{8-1000\mu m}\sim 10^{10}L_{\odot}$) in order to link the well-studied ultra luminous infrared galaxies to normal star-forming galaxies. Our findings are as follows: 
(i) we found that more IR luminous galaxies tend to have smaller local galaxy density, being consistent with the picture where luminous IRGs are created by the merger-interaction of galaxies that happens more often in lower density regions;
(ii)  the fractions of AGNs increase as a function of $L_{ir}$;
(iii) there is a good correlation between $L_{ir}$ and $L_{[OIII]}$ for AGNs, suggesting both of the parameters can be a good estimator of the total power of AGNs;
 (iv) a good correlation is found between $L_{ir}$ and optically estimated star formation rate (SFR) for star-forming galaxies, suggesting  $L_{ir}$ is a good indicator of galaxy SFR. However, caution is needed when SFR is estimated using $L_{ir}$, i.e., high SFR galaxies selected by $L_{ir}$ is frequently merger/interaction, whereas high SFR galaxies selected by optical emission is often normal spiral or Magellanic-cloud like irregular galaxies. 
(v) more IR luminous galaxies have slightly larger $H\alpha/H\beta$ ratio;
(vi) more  IR luminous galaxies have more centrally-concentrated morphology, being consistent with the morphological appearance of galaxy-galaxy merger remnants. Optical image of ultra/very luminous infrared galaxies also show frequent signs of merger/interaction. 
(vii) comparison with the SED synthesis models indicate that majority of luminous infrared galaxies ($L_{8-1000\mu m} > 10^{11}L_{\odot}$) may be in a post-starburst phase, sharing a similar (but not the same) merger/interaction origin with E+A galaxies. 

\end{abstract}

\section{Introduction}\label{intro}

 One of the most important achievements of the infrared astronomical satellite (IRAS; Neugebauer et al. 1984) is the discovery of a significant population of luminous infrared galaxies (Soifer et al. 1984; Sanders \& Mirabel 1996; Sanders 2003). Especially, the most energetic objects, the ultra luminous infrared galaxies, has been a subject of extensive debate (ULIRGs; $L_{8-1000\mu m}>10^{12}L_{\odot}$; e.g., Sanders et al. 1988; Borne et al. 2000; Kim et al. 2002) since ULIRG's  bolometric luminosity is comparable to that of QSOs (Soifer et al. 1987) and dominate the top end of the galaxy luminosity function.  The high infrared (IR) luminosity is attributed to dust emission, with the heating source being varying combinations of starbursts (Genzel et al. 1998; Lutz et al. 1998) and active galactic nuclei (AGNs; Veilleux et al. 1999a,b;Nagar et al. 2003). The luminosity functions of infrared galaxies show that at $L_{ir}>10^{11}L_{\odot}$, infrared selected galaxies become more numerous than optically selected starburst and Seyfert galaxies of comparable bolometric luminosity (Soifer et al. 1987). At  $L_{ir}>10^{12}L_{\odot}$, ULIRGs exceed the space densities of QSOs by a factor of 1.5-2 (Sanders et al. 1989).
Some investigators have proposed that ULIRGs may represent the initial, dust-enshrouded stages of QSOs (e.g., Sanders et al.1999). Some ULIRGs (especially cool ULIRGs) also appear to be forming moderately massive ($L^*$) field elliptical galaxies (e.g., Melnick \& Mirabel 1990;Kormendy \& Sanders 1992; Genzel et al. 2001). Their near-IR light distributions often fits an $r^{1/4}$-law (Wright et al. 1990; Stanford \& Bushouse 1991; Doyon et al. 1994; Scoville et al. 2000; Veilleux et al. 2002). 
Based on both optical and infrared morphology, many studies found that the majority of ULIRGs are strongly interacting or advanced merger systems (e.g. Borne et al. 2000).

On the other hand, lower luminosity infrared galaxies ($10^{9}L_{\odot}<L_{ir}<10^{12}L_{\odot}$) have not been the subject of much scrutiny (however, see Kim et al. 1995; Wu et al. 1998a,b; Arribas et al. 2004). However, lower luminosity infrared galaxies are also an interesting class of galaxies for various reasons: it has been revealed that the properties of ULIRGs appear to correlate with IR luminosity (Sanders \& Mirabel 1996). Therefore, in order to understand infrared galaxies (IRGs) as a whole, we need to study IRGs at a lower luminosity range. The luminosity of lower luminosity IRGs investigated here are between ULIRGs and more popular spiral galaxies ($10^{9}L_{\odot}<L_{ir}<10^{12}L_{\odot}$). Therefore, lower luminosity IRGs might provide  a link between normal spiral galaxies and infrared luminous galaxies.  In addition, lower luminosity IRGs are much more frequent than ULIRGs, and thus, it is important to study VLIRGs in order to understand cosmological evolution of galaxies as a whole.   

However, previously, it has been difficult to study statistical number of lower luminosity IRGs. It has been simply difficult to obtain telescope time to follow-up thousands of IRGs both spectroscopically and photometrically. 
   One of the pioneers is Arribas et al. (2004) who observed 19 very luminous infrared systems (VLIRGs; $10^{11}L_{\odot}<L_{ir}<10^{12}L_{\odot}$) in order to probe an important link between ULIRGs and VLIRGs. They found that the photometric properties of ULIRG and VLIRG samples, considered as a whole, was indistinguishable at optical wavelengths. 
   Wu et al. (1998a,b) carried out spectroscopic observation of 73 VLIRGs to find that 60\% of VLIRGs exhibit AGN-like spectra, and that 56\% of VLIRGs show strong merging/interaction signatures. The fractions are smaller, but these are similar characteristics to ULIRGs.
 With the advent of the Sloan Digital Sky Survey (SDSS), we have a chance to step forward. By matching up $\sim$360,000 spectroscopic galaxies in the SDSS Data Release 3 (SDSS DR3; Abajazian et al. 2004) with the IRAS data,  we have obtained the opportunity to study optical photometric and spectroscopic properties of unprecedented number of 4248 infrared galaxies, including three hyper luminous infrared galaxies and 178 ultra luminous infrared galaxies.   Unless otherwise stated, we adopt the best-fit WMAP cosmology: $(h,\Omega_m,\Omega_L) = (0.71,0.27,0.73)$ (Bennett et al. 2003).

\section{Data}\label{data}
\subsection{Matching up the IRAS and the SDSS catalogs}
  
 In order to create a large sample of galaxies with both optically measured redshift and mid- and far-infrared photometry, we have positionally matched up the IRAS galaxies (IRAS faint source catalog Version 2; Moshir et al. 1992; FSC92 hereafter) and the SDSS galaxies (The Sloan Digital Sky Survey Data Release 3; Abazajian et al. 2004) with spectroscopically measured redshifts. Moshir et al. (1992) provided more accurate fluxes and positions than that of 1988 and is considered as the standard reference catalog for IRAS sources (However, note possible inaccuracies in using catalog line fluxes).  From IRAS catalog, we exclude IRAS sources with $f_{60}<f_{12}$ to remove contamination from infrared bright stars which are known to peak around 12 $\mu$m (Cohen et al. 1987). This criterion has been known to be efficient  in selecting galaxies, while excluding most Galactic objects. ULIRGs typically have $f_{60}/f_{12}>20$. 
    Positional accuracy of the IRAS catalog in cross-scan direction is $\sim$10 arcsec. Therefore, if an IRAS galaxy has a counterpart in the SDSS within 20 arcsec, we regard it a match. Total number of galaxies with both the IRAS and the SDSS detections is 4248, which is one of the largest data-sets with both optical redshift and the IR photometry.  When we construct a volume limited sample, we select a redshift and absolute magnitude range of $z\leq 0.06$ and $M_r \leq -19.42$ to maximize the number of galaxies in the volume limited sample, resulting in 1666 galaxies in the sample.

\begin{figure}
\begin{center}
\includegraphics[scale=0.4]{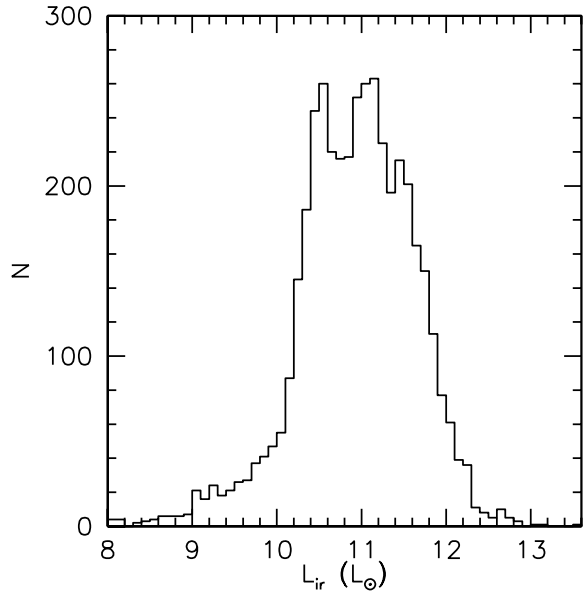}
\end{center}
\caption{The histogram of total infrared luminosity, ($L_{ir}$).  
}\label{fig:Lir_hist} 
\end{figure}

\subsection{Infrared Luminosity}

\begin{figure*}
\begin{center}
\includegraphics[scale=0.81]{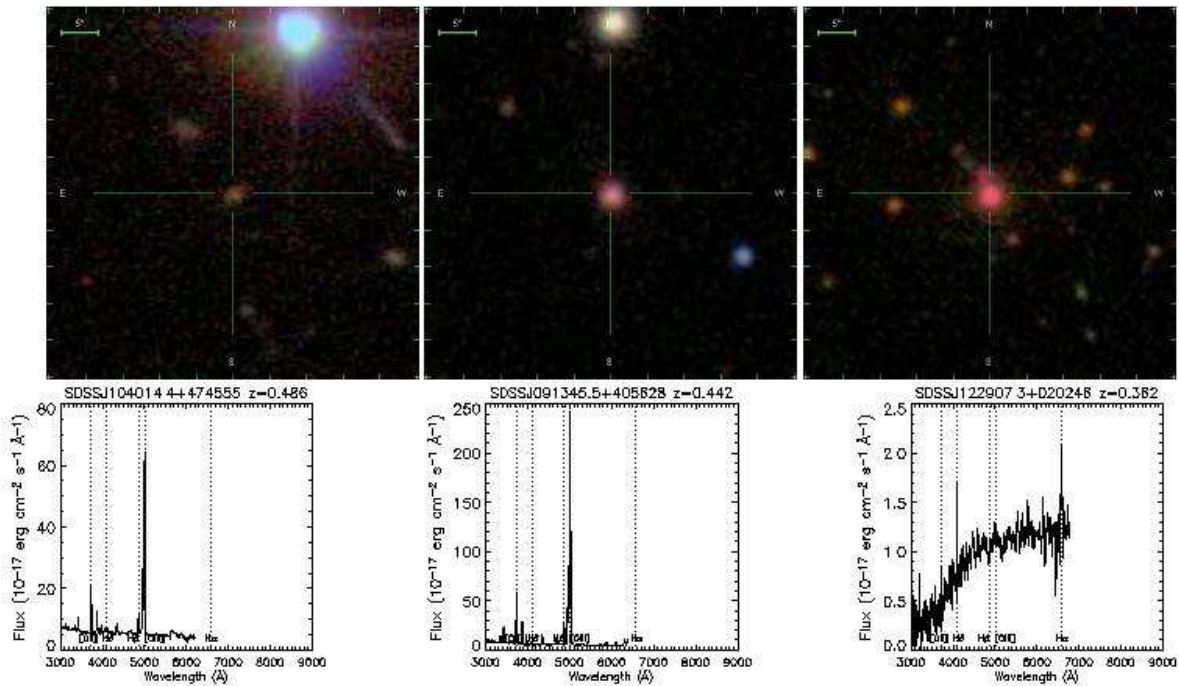}
\end{center}
\caption{ Three HLIRGs ($10^{13.0}L_{\odot}<L_{ir}$) identified in this work. The upper panel shows $g,r,i$-composite images. The corresponding spectra with name and redshift are presented in the lower panel. Their luminosity is $log (L_{ir})=$13.05,13.57 and 13.19$L_{\odot}$, respectively.
}\label{fig:13_images}
\end{figure*}

\begin{figure*}
\begin{center}
\includegraphics[scale=0.81]{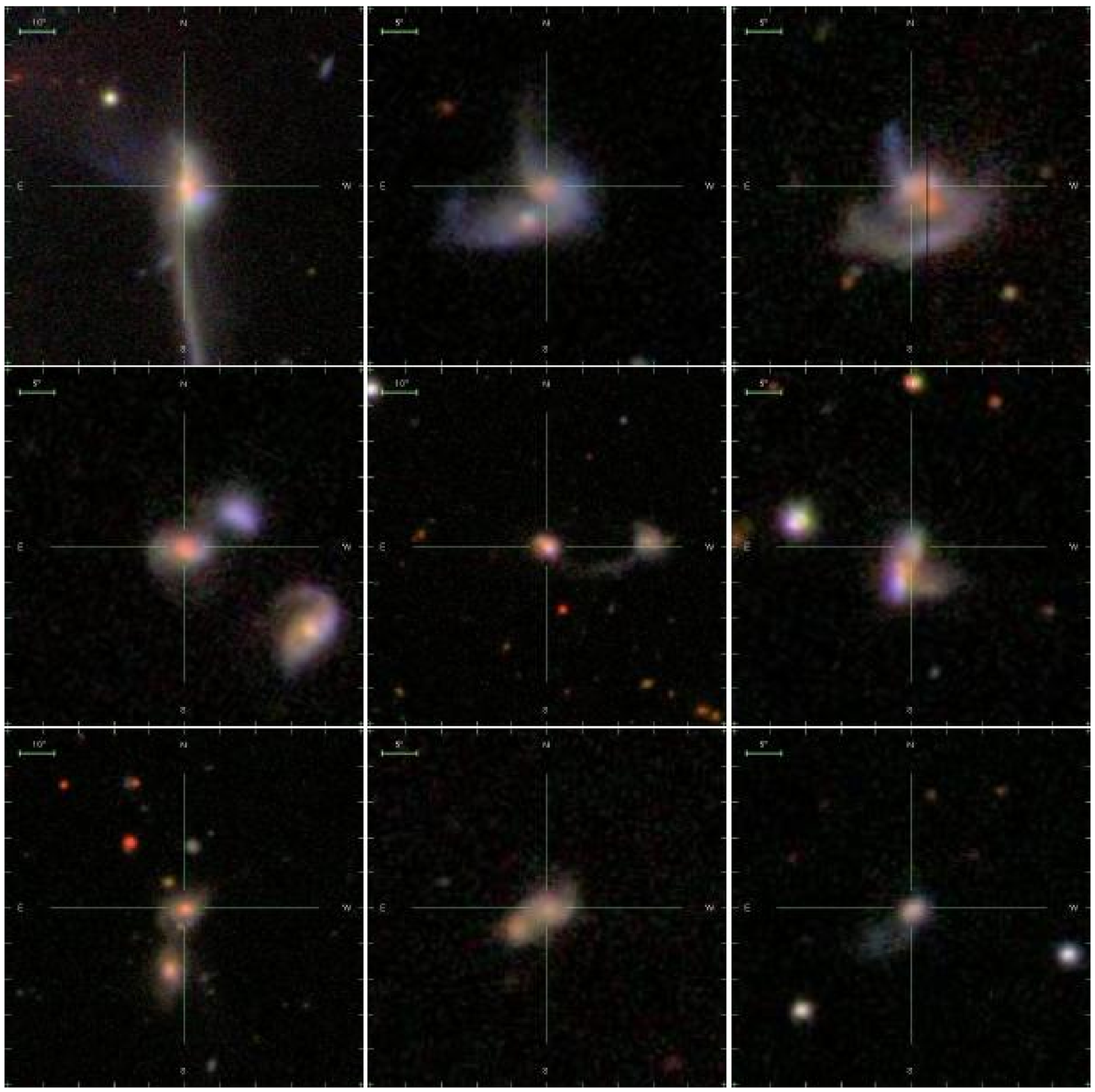}
\end{center}
\caption{Examples of $g,r,i$-composite images of ULIRGs ($10^{12.0}L_{\odot}<L_{ir}\leq 10^{13.0}L_{\odot}$). The images are sorted from low to high redshift. Only 9 lowest redshift galaxies are shown. The corresponding spectra with name and redshift are presented in Fig.\ref{fig:12_spectra}. (The spectrum of the same galaxy can be found in the same column/row panel of Fig.\ref{fig:12_spectra}.)
}\label{fig:12_images}
\end{figure*}

\begin{figure*}
\begin{center}
\includegraphics[scale=0.81]{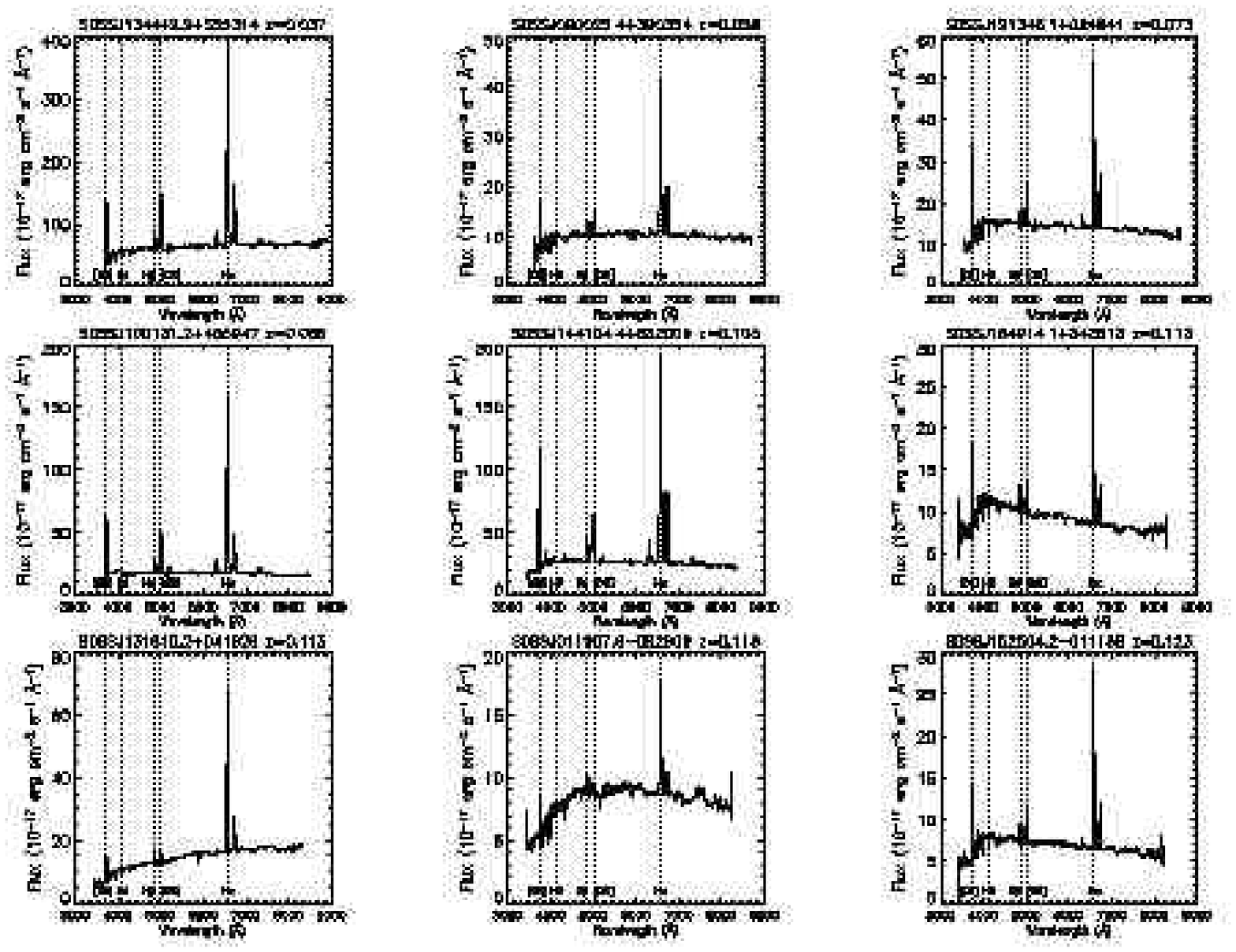}
\end{center}
\caption{Example spectra of 9 lowest redshift ULIRGs. The spectra are sorted from low redshift. Each spectrum is shifted to the restframe wavelength and smoothed using a 20 \AA\ box. The corresponding images are shown in Fig. \ref{fig:12_images}. (The image of the same galaxy can be found in the same column/row panel of Fig.\ref{fig:12_images}).
}\label{fig:12_spectra}
\end{figure*}

\begin{figure}
\begin{center}
\includegraphics[scale=0.4]{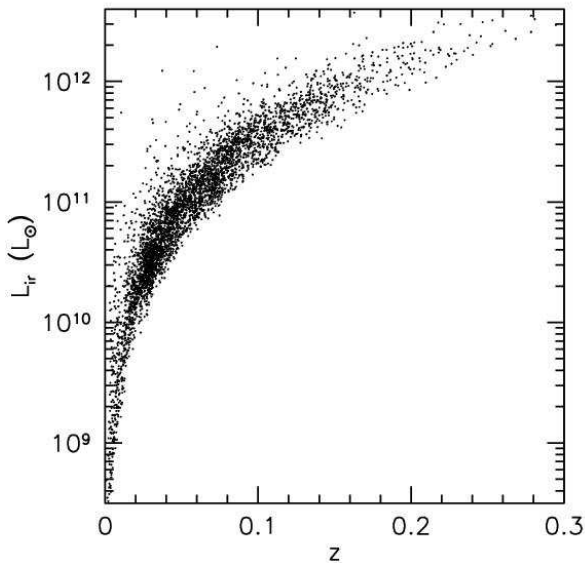}
\end{center}
\caption{ Total infrared luminosity measured with the IRAS, ($L_{ir}$), is plotted against redshift measured with the SDSS.
}\label{fig:z_lir} 
\end{figure}

 For each galaxy in the matched catalog, we compute infrared flux using the following equation. 
\begin{equation}
   F_{ir}=1.8 \times (13.48 \times f_{12}+ 5.16 \times f_{25} +2.58 \times f_{60} +1.0\times f_{100})
\label{eq:fir}
 \end{equation}
The IRAS fluxes are taken directly from the FSC92. Since many of the high luminosity galaxies emit a significant portion of their infrared luminosity shortward of 40 $\mu$m, this expression provides a significantly better determination of the total infrared flux than the more commonly used $F_{ir}$ determined by fitting a single temperature dust model to the 60 $\mu$m and 100 $\mu$m fluxes (Kim \& Sanders 1998). This flux, $F_{ir}$, is then converted to luminosity ($L_{ir}$) using the redshift measured by the SDSS. To keep consistency with previous work, we use $h=0.75$ and $q_0=0$ only for this computation. We show the histogram of total infrared luminosity ($L_{ir}$) in Figure \ref{fig:Lir_hist}. We have identified three hyper luminous infrared galaxies (HLIRG; $10^{13.0}L_{\odot}< L_{ir}$) and show their images and spectra in Figure \ref{fig:13_images}. The most luminous object in the sample is SDSSJ091345.5+405628 with $L_{ir}=10^{13.57}L_{\odot}$. Our sample also includes 178 ULIRGs, which is one of the largest of the kind. We show 9 lowest redshift ULIRGs in Figure \ref{fig:12_images}. The corresponding optical spectra are shown in Figure \ref{fig:12_spectra}.
In Figure \ref{fig:z_lir}, $L_{ir}$ is plotted against redshift. In this work, we call galaxies with $10^{13.0}L_{\odot}< L_{ir}$ as hyper luminous infrared galaxies (HLIRGs), galaxies with $10^{12.0}L_{\odot}< L_{ir}\leq 10^{13}L_{\odot}$ as ultra luminous infrared galaxies (ULIRGs), galaxies with $10^{11.0}L_{\odot}< L_{ir}\leq 10^{12}L_{\odot}$ as very luminous infrared galaxies (VLIRGs), galaxies with  $10^{10.5}L_{\odot}< L_{ir}\leq 10^{11.0}L_{\odot}$ as moderately luminous infrared galaxies (MLIRGs),and $10^{10.0}L_{\odot}< L_{ir}\leq 10^{10.5}L_{\odot}$ as lower luminosity infrared galaxies (LLIRGs), respectively. When we mention infrared galaxies (IRGs), the sample includes all the classes of infrared galaxies plus galaxies with $L_{ir}\leq 10^{10.0}L_{\odot}$.

\section{Results}\label{sec:Results}

\subsection{The environment of Infrared Luminous Galaxies}\label{sec:Mpc}

%
 
In this subsection, we investigate the environment of luminous infrared galaxies.
 It has been often suggested that the ULIRGs are caused by the major merger of multiple galaxies (Borne et al. 2000; Taniguchi \& Shioya 1998). However, these results are based on the eye-based scrutiny of the image/morphology of the central IRGs, often with disturbed/dynamical tail signatures. As an independent way of measuring the environment of IRGs, we study local galaxy density distribution of IRGs.

\begin{figure*}
\begin{center}
\includegraphics[scale=0.4]{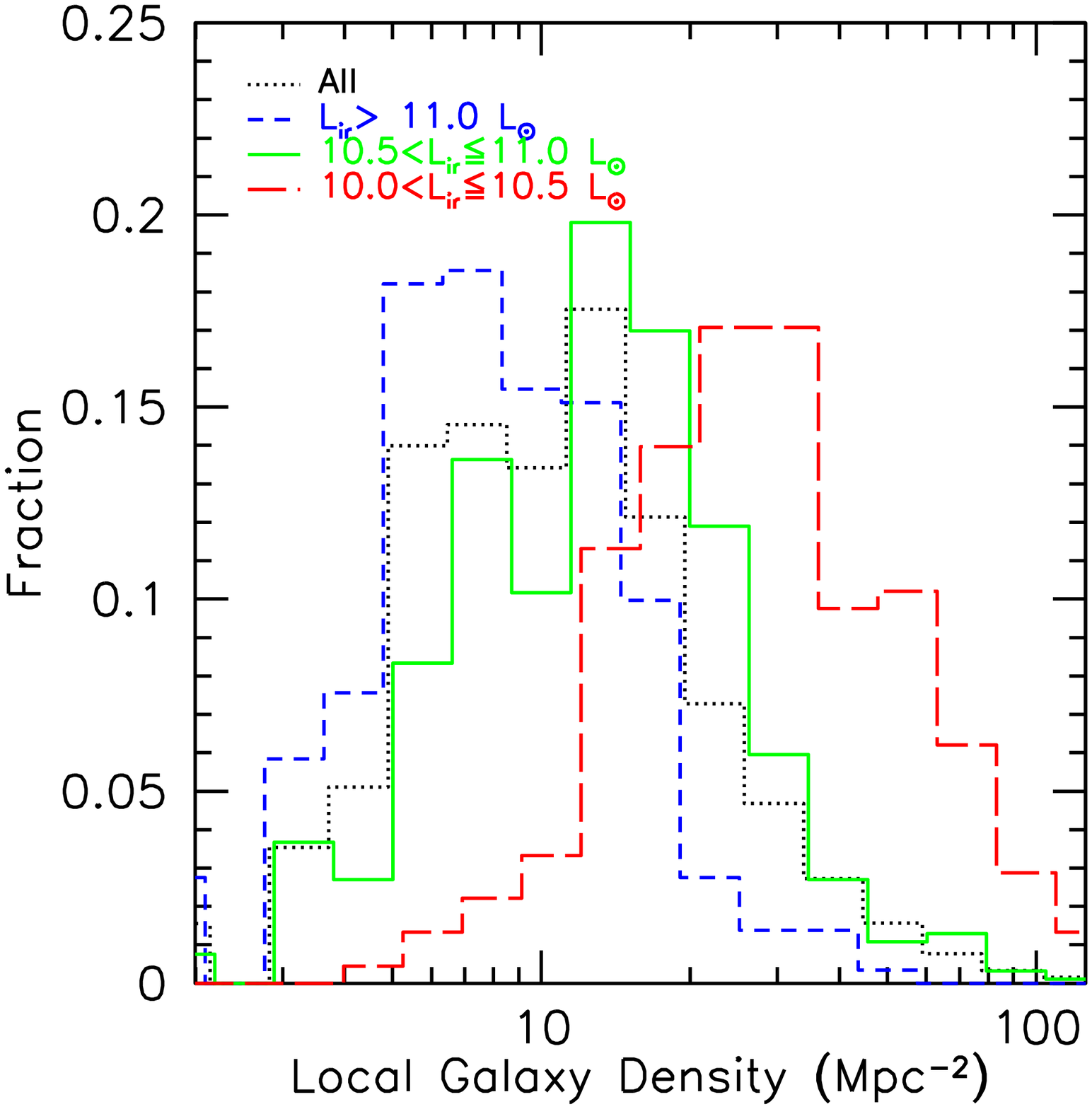}
\includegraphics[scale=0.4]{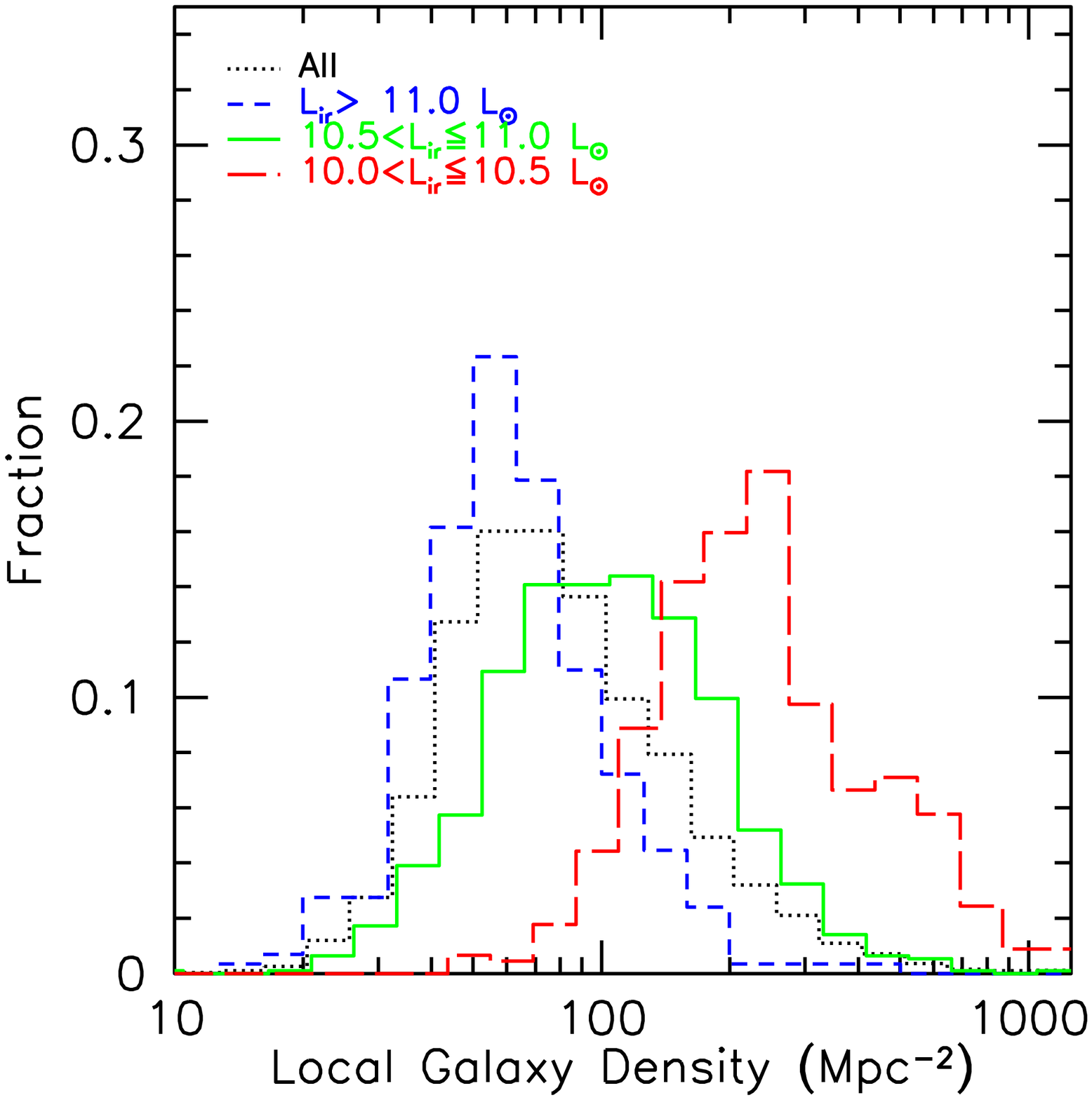}
\includegraphics[scale=0.4]{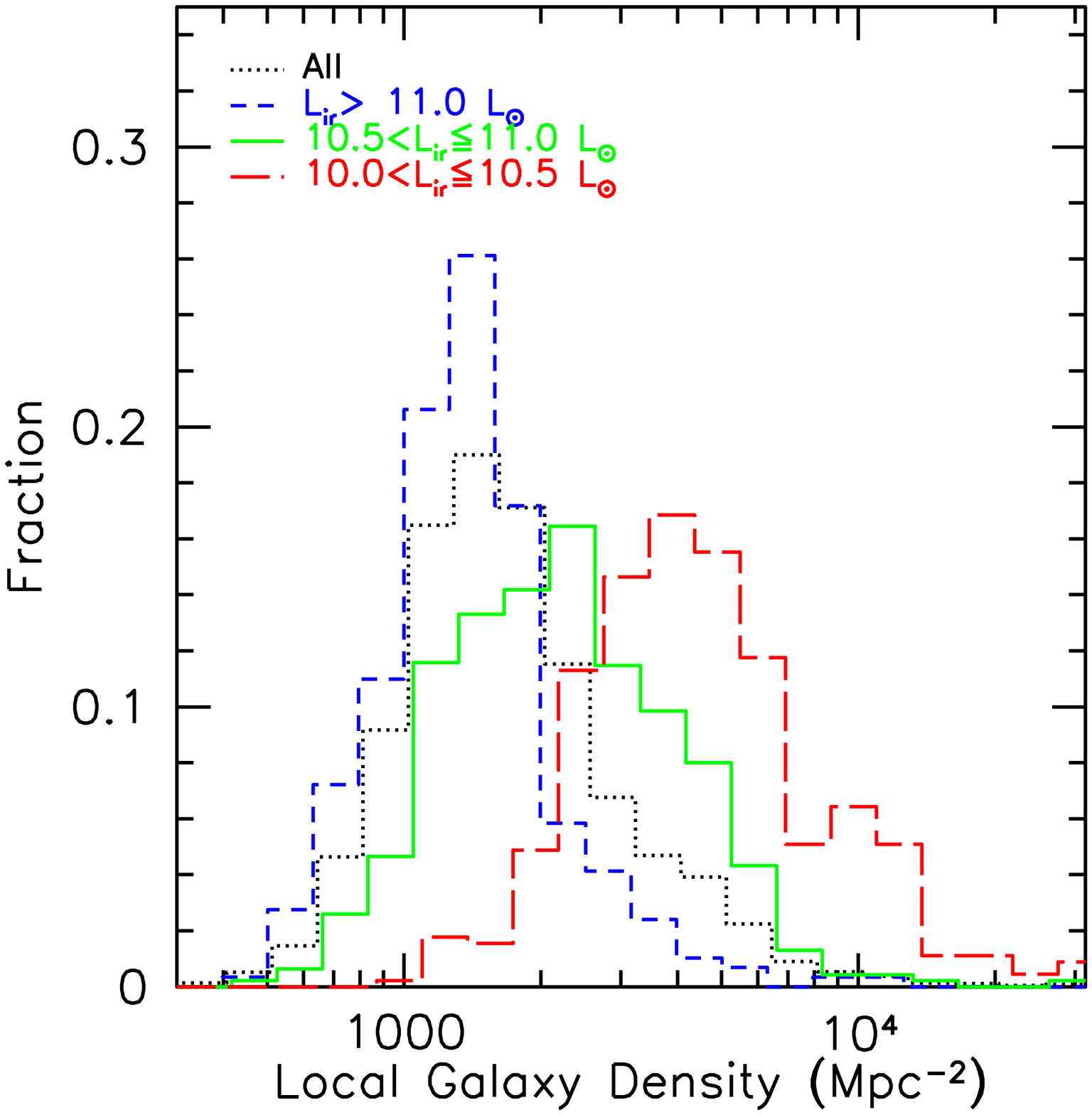}
\includegraphics[scale=0.4]{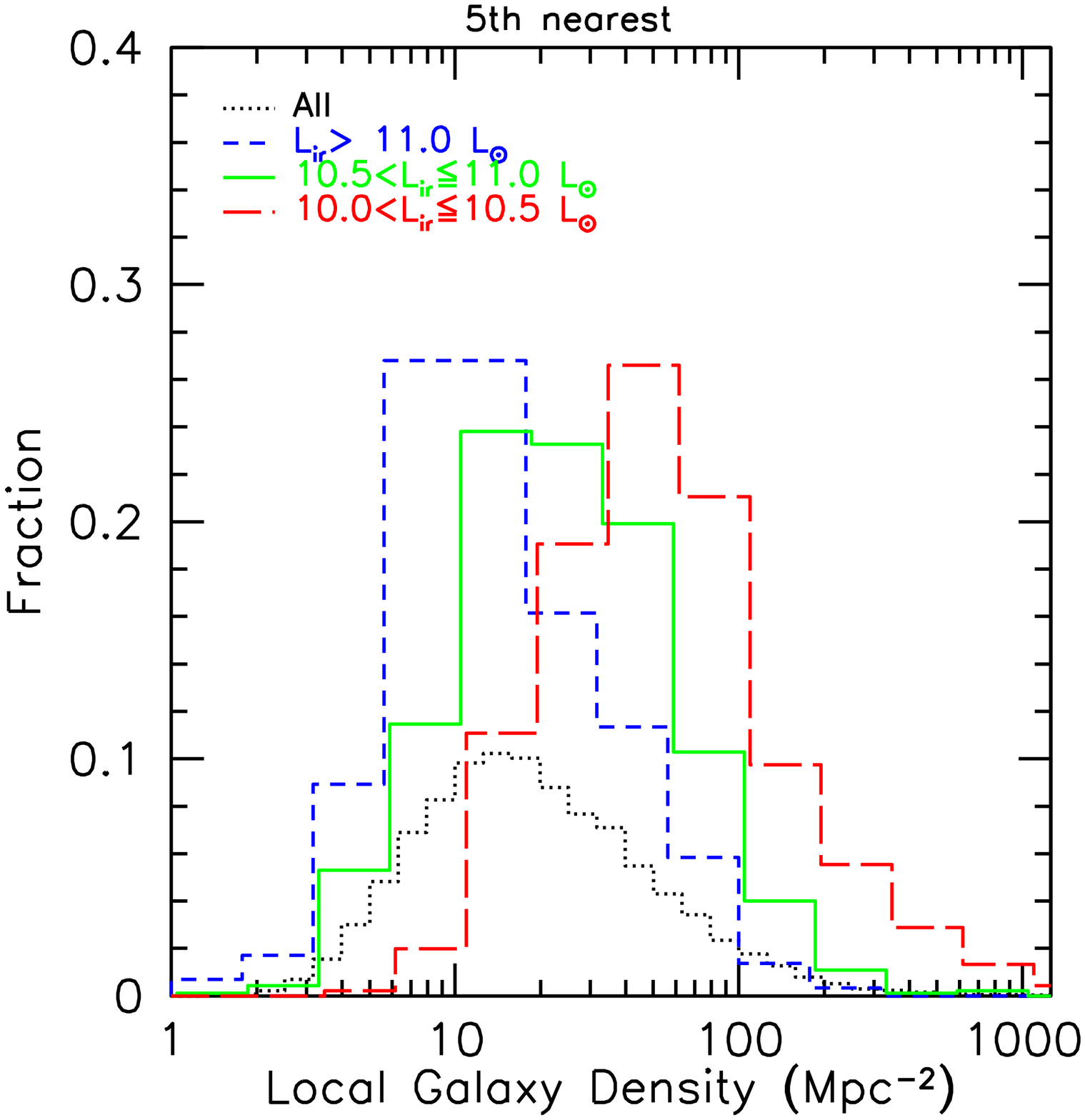}
\end{center}
\caption{
Normalized distributions of local galaxy density. The local galaxy densities are measured in 0.5 Mpc (top left panel), 1.5 Mpc (top right panel) and 8 Mpc (bottom left panel) of radius in angular direction and $\pm$1000 km s$^{-1}$ in the line of sight direction within the volume limited sample ($z\leq 0.06$ and $Mr\leq -19.42$). In the bottom right panel, the density is measured as a distance to the 5th nearest galaxy in angular direction and $\pm$1000 km s$^{-1}$ in the line of sight direction within the volume limited sample. The dotted, short-dashed, solid, long-dashed lines denote all galaxies in the volume limited sample (regardless of IRAS detection), VLIRGs, MLIRGs, and LLIRGs, respectively. ULIRGs and HLIRGs are not plotted due to their small number in the volume limited sample.
 }\label{fig:density}
\end{figure*}

 We use the volume limited sample ($z\leq 0.06$ and $M_r \leq -19.42$) in this subsection to ensure neighbors are measured in the same absolute magnitude range for each and every galaxy in the sample.  First, we measure local galaxy density in a fixed physical scale. In Figure \ref{fig:density}, the local galaxy densities are measured in 0.5 Mpc (top left panel), 1.5  Mpc (top right panel) and 8 Mpc (bottom left panel) of radius in angular direction and $\pm$1000 km s$^{-1}$ in the line of sight direction within the volume limited sample ($z\leq 0.06$ and $Mr\leq -19.42$). Number of galaxies in each columnar volume is divided by the physical surface area to be converted to local galaxy density. The density computed in a fixed physical scale allows us to investigate the environment of IRGs at a specific scale.
In contrast, in the bottom right panel of Figure \ref{fig:density}, the density is measured using the distance to the 5th nearest galaxy in angular direction and $\pm$1000 km s$^{-1}$ in the line of sight direction within the volume limited sample. Since the distance to the 5th nearest scales with galaxy density, this density provides us with a scale-free estimate of the environment of IRGs, spanning a much wider density range of 4 orders. This parameter is often used in previous studies (e.g., Goto et al.2003b,c).  In  Figure \ref{fig:density}, the dotted, short-dashed, solid, long-dashed lines denote all galaxies in the volume limited sample (regardless of IRAS detection), galaxies with VLIRGs, MLIRGs and LLIRGs, respectively. ULIRGs are not included in this analysis due to their small number in the volume limited sample. 

It is immediately noted that more infrared luminous galaxies have smaller local galaxy density at all scales. For example, in case of the 5th nearest density (lower right panel), median density is 1.11, 1.33, and 1.68 Mpc$^{-2}$ for VLIRGs, MLIRGs and LLIRGs, respectively. Compared with the density distribution of all galaxies in the volume limited sample (dotted line; we regard this as the field sample), VLIRGs have significantly lower density, and MLIRGs and LLIRGs have higher density than the field sample. This is a contrasting result compared with that from optical luminosity. In optical, more luminous galaxies tend to exist in higher density regions --- often the most luminous ones sit at the center of galaxy clusters (e.g., Goto et al. 2002a,b). Perhaps, the total infrared luminosity ($L_{ir}$) reflects star formation rate (SFR) of galaxies rather than the total mass in galaxies. Then, this trend can be understood as more star-forming galaxies reside in lower density region. If the origin of VLIRGs and MLIRGs is merger/interaction of multiple galaxies, it is also naturally explained that they preferentially exist in lower density regions, where relative velocity of galaxies are small, and thus, dynamical merging of galaxies happens more frequently. 
 It also should be noted that IRGs do not have density peaked at the highest density, suggesting majority of IRGs are not in galaxy clusters, but in the field region.

  Using the same volume limited samples, we compute median $M_r$ of each sample in Table \ref{tab:abr}. We use Blanton et al. (2003, v3\_2) for k-correction. As is expected from the $L_{ir}$, median $M_r$ is brighter for the sample with larger $L_{ir}$. However, the increase in $r$-band flux is only by a factor of $\sim$1.5, compared to the order of $\sim 2$ increase in $L_{ir}$. The $M_r$ of MLIRGs and LLIRGs are even comparable to that of all galaxies in the volume limited sample.

\begin{table}
\caption{
Median absolute magnitudes in the SDSS $r$-band ($M_r$) within the volume limited sample ($z\leq 0.06$ and $Mr\leq -19.42$).
}\label{tab:abr}
\begin{center}
\begin{tabular}{crr}
\hline
 Sample   & Median $M_r$ & $N_{galaxies}$\\
\hline
\hline
ALL& $-$20.10  & 20753\\ 
VLIRGs & $-$20.61 & 289\\ 
MLIRGs & $-$20.34 & 924\\ 
LLIRGs & $-$20.05 & 451\\ 
\hline
\end{tabular}
\end{center}
\end{table}


\subsection{Starburst/AGN classification}

\begin{figure}
\begin{center}
\includegraphics[scale=0.4]{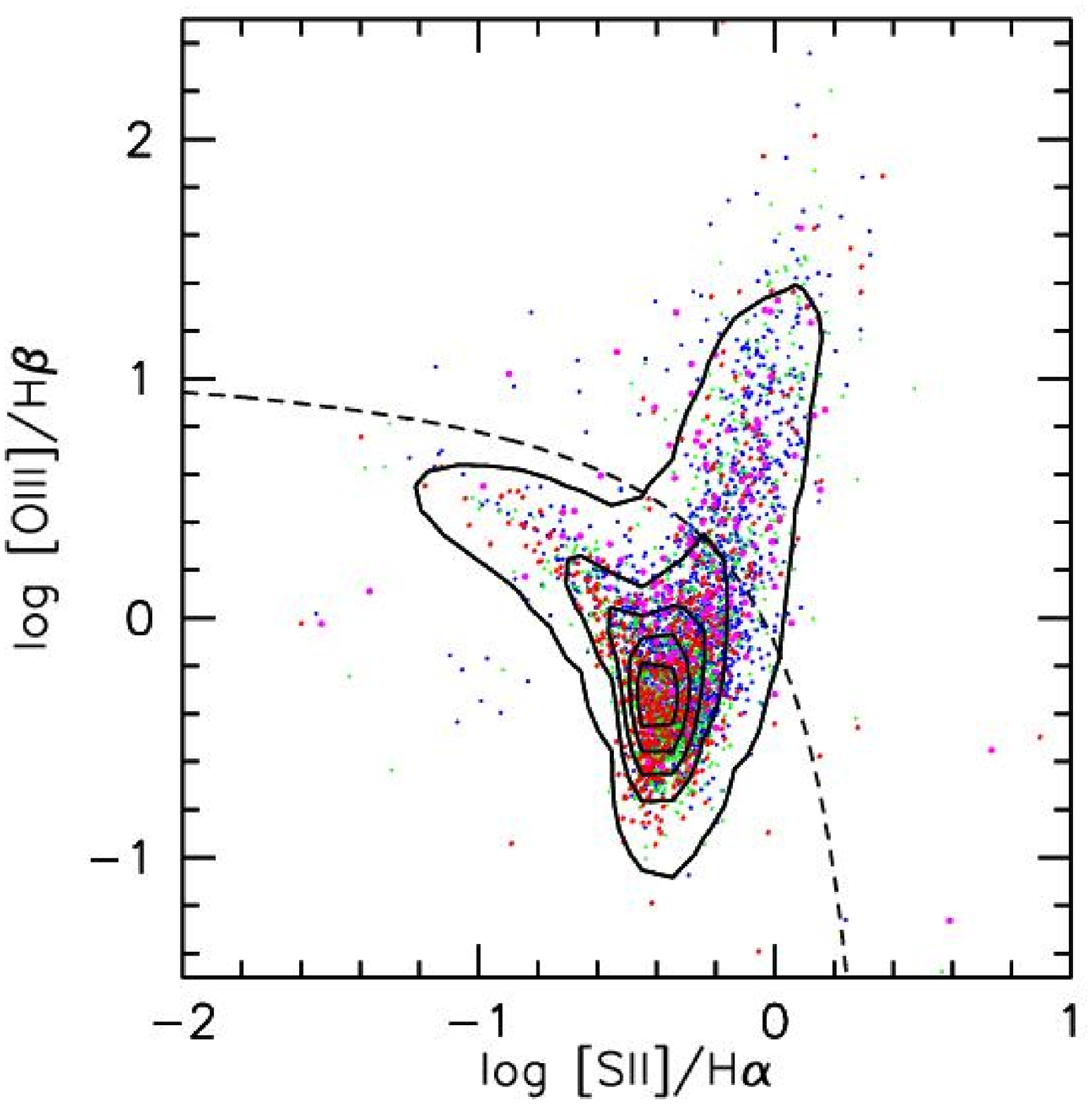}
\end{center}
\caption{Emission line ratios used to select AGNs from our sample. The contour shows distribution of all galaxies in the SDSS with $r<17.77$ (regardless of IR detection). The dashed line is the criterion between starbursts and AGNs described in Kewley et al. (2001). Galaxies with line ratios higher than the dashed line are regarded as AGNs. The magenta, blue, green, and red, dots are for ULIRGs, VLIRGs, MLIRGs, and LLIRGs, respectively.
}\label{fig:Osterbrook} 
\end{figure}

In the following subsections, we investigate optical properties of IRGs in detail. However, there are two types of energy sources in infrared radiation: starburst and active galactic nuclei (AGN); it is critical to separate these two in order to understand physics behind infrared radiation. 
In this work, we separate AGNs from starburst galaxies using the optical emission line ratios as described in Kewley et al. (2001). Figure \ref{fig:Osterbrook} shows emission line ratios of [OIII](5008\AA)/H$\beta$ versus [NII](6585\AA)/H$\alpha$.  Here, we use the emission line flux measured by the SDSS pipeline (Stoughton et al. 2002).  The contour shows distribution of all galaxies in the SDSS DR3 with $r<17.77$ (regardless of IR detection). The magenta, blue, green, and red dots are for ULIRGs, VLIRGs, MLIRGs, and LLIRGs, respectively. The dotted line is the criterion used to divide AGNs and starbursts proposed by Kewley et al. (2001).  We take a conservative approach in selecting AGNs. We only regard a galaxy as an AGN when all four lines are detected in emission and the line ratios are above the dotted line in Fig.\ref{fig:Osterbrook}. The rest of the galaxies (below the dotted lines or with unmeasurable emission lines) is regarded as starburst galaxies. 
 In our sample, there are 810 AGNs and 3438 star-forming galaxies with the IRAS detections. Both of these are one of the largest number of IRGs observed both in optical and IR.

\begin{figure}
\begin{center}
\includegraphics[scale=0.4]{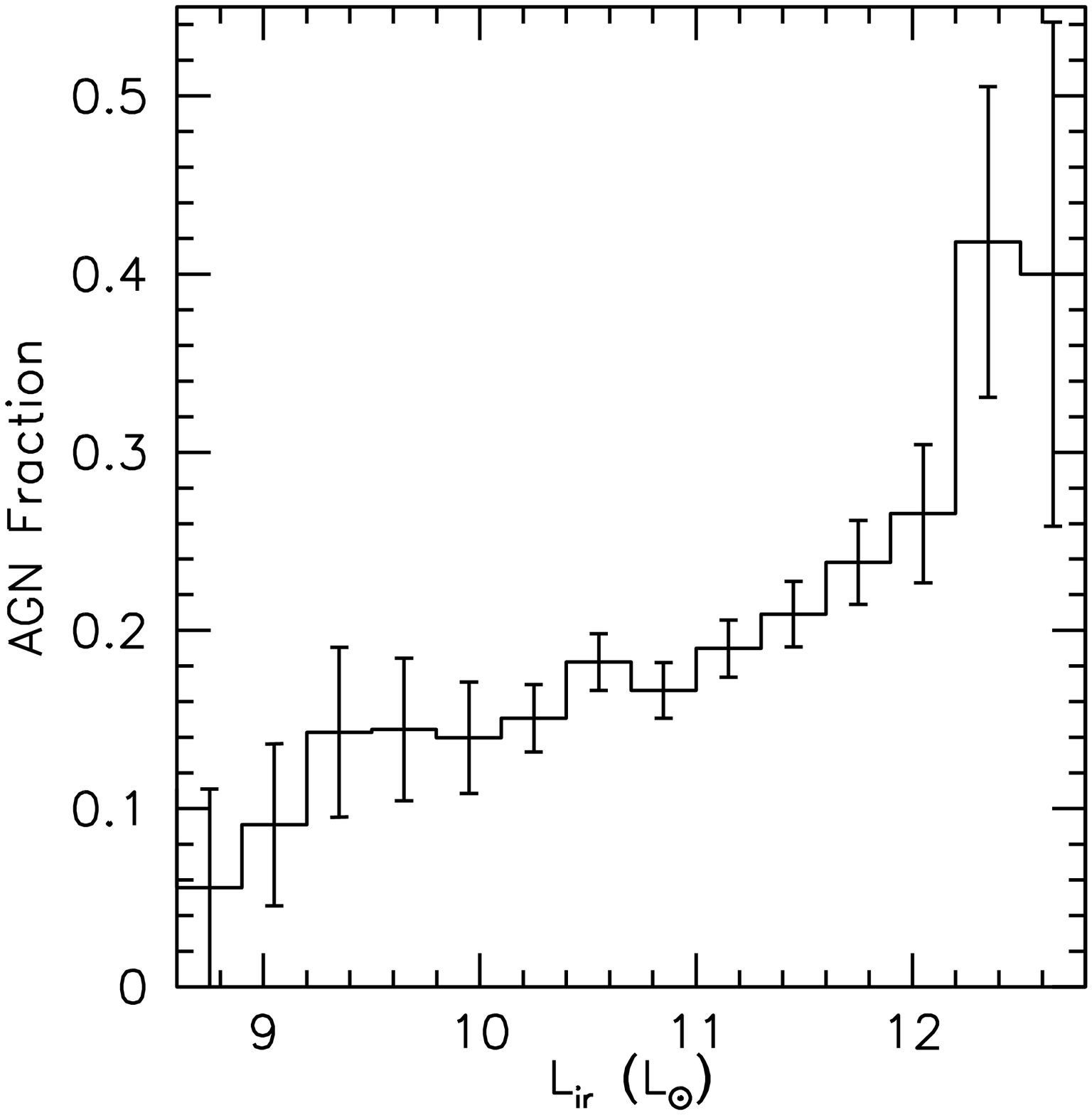}
\end{center}
\caption{ Fractions of AGNs as a function of $L_{ir}$. The error bars are based on Poisson statistics. AGNs are selected using the optical emission line ratios in Fig.\ref{fig:Osterbrook}.
}\label{fig:agn_frac} 
\end{figure}

\begin{figure}
\begin{center}
\includegraphics[scale=0.4]{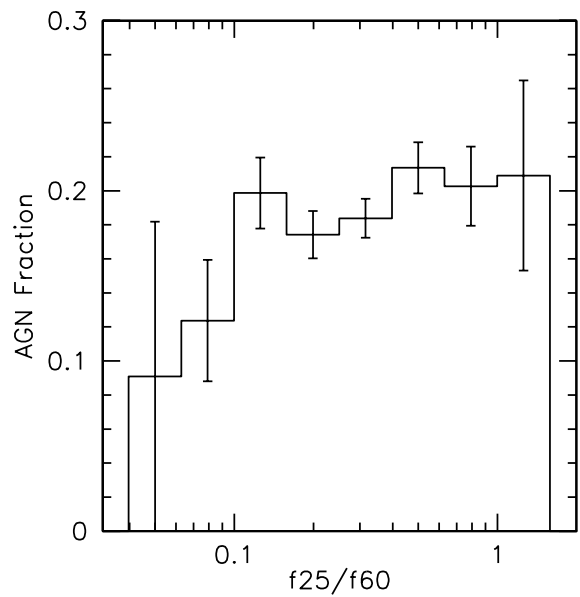}
\end{center}
\caption{
  Fractions of AGNs as a function of $f25/f60$ IR color. The error bars are based on Poisson statistics. AGNs are selected using the optical emission line ratios in Fig.\ref{fig:Osterbrook}.
  }\label{fig:agn_frac_ircolor}
\end{figure}

In Figure \ref{fig:Osterbrook}, it is also found that more ULIRGs and VLIRGs (magenta and blue) are above the dotted line (AGN/starburst criterion), suggesting more ULIRGs/VLIRGs are AGNs than lower luminosity IRGs. To clarify, in Figure \ref{fig:agn_frac}, we plot fractions of AGNs selected in Figure \ref{fig:Osterbrook} as a function of total infrared luminosity. It is understood that at a higher luminosity, more galaxies are powered by AGNs, especially at $L_{ir}>10^{12}L_{\odot}$. In Figure \ref{fig:agn_frac_ircolor}, we show fractions of AGNs as a function of $f25/f60$ IR color. As previously suggested (e.g., Veilleux et al. 1999), at $f25/f60<0.2$, the fractions of AGNs decreases, whereas at $f25/f60\geq 0.2$ the fractions of AGNs are almost constant. 

 Previously, it has been suggested that IR galaxies have lower excitation than optically selected starburst galaxies (Sarazin 1976; Allen et al. 1985; Leech et al. 1989). However, in Figure \ref{fig:Osterbrook}, the distributions of dots (IRGs) are consistent with that of the contours (all SDSS galaxies). Perhaps, the comparison of excitation ratio depends on how you select a starburst sample. Our data show that when compared with all the SDSS galaxies, the excitation of IRGs is not significantly lower than that of the optically selected galaxies.
                                                                                                                                                                                                                                                                                                                                                                                                                        

\subsection{Optical Line Properties of Infrared Luminous AGNs}

\begin{figure}
\begin{center}
\includegraphics[scale=0.4]{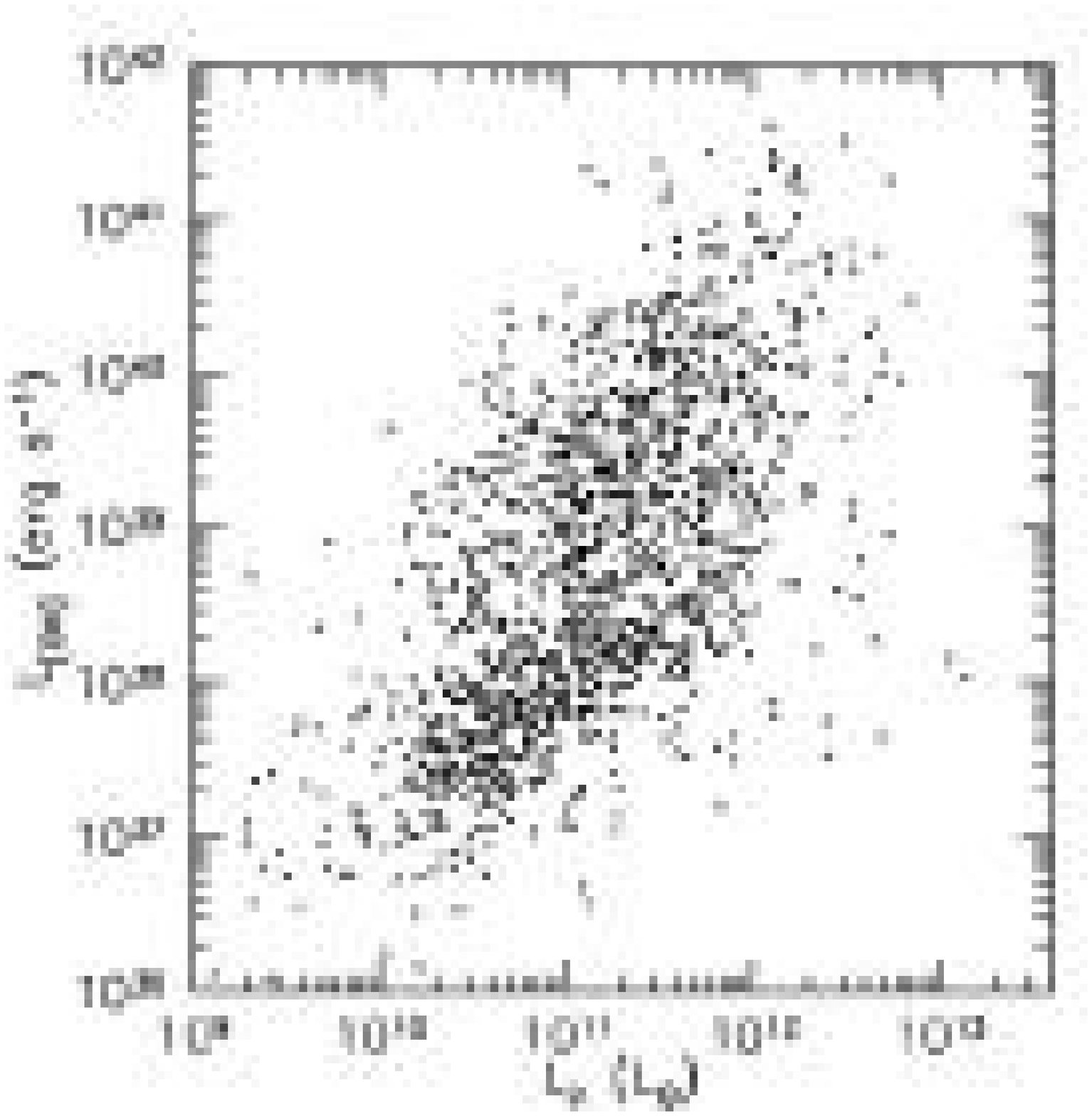}
\end{center}
\caption{ Luminosity of the [OIII](5008\AA) emission line is plotted against total infrared luminosity for AGNs.
}\label{fig:oiii_lir} 
\end{figure}

\begin{figure}
\begin{center}
\includegraphics[scale=0.4]{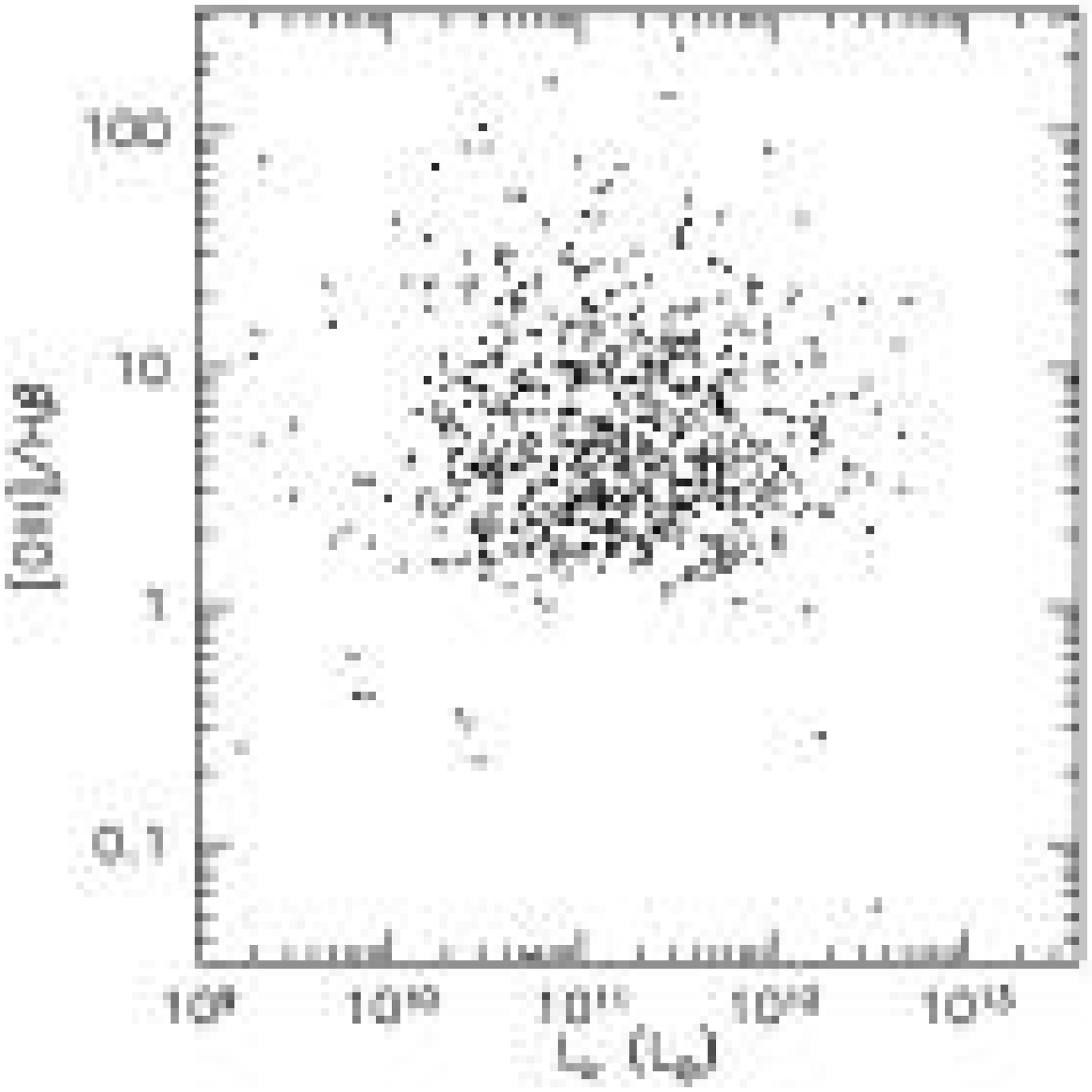}
\includegraphics[scale=0.4]{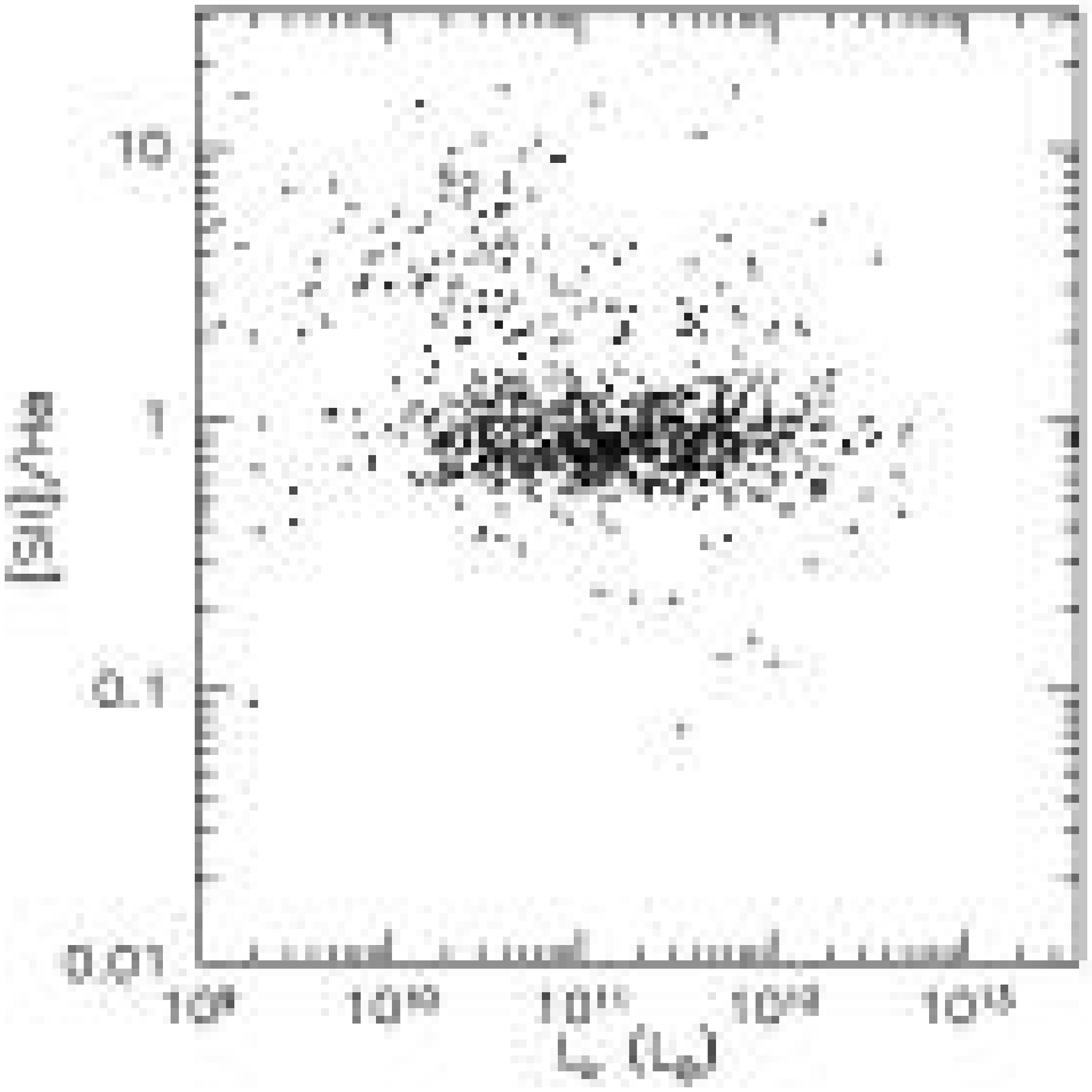}
\end{center}
\caption{Excitation line ratio of AGNs as a function of $L_{ir}$. The upper panel shows [OIII]/H$\beta$ ratio as a function of  $L_{ir}$. The lower panel is for [SII]/H$\alpha$ ratio against $L_{ir}$.}\label{fig:agn_excite}
\end{figure}

 In this subsection, we investigate optical emission line properties of IR luminous AGNs. In Figure \ref{fig:oiii_lir}, we plot luminosity of the [OIII](5008\AA) emission line against total infrared luminosity for AGNs selected in Figure \ref{fig:Osterbrook}. It is immediately noticed that there is a good correlation between $L_{[OIII]}$ and $L_{ir}$, suggesting both $L_{[OIII]}$ and $L_{ir}$ can be a good estimator of the total power of the AGNs. 
 On the other hand, the excitation line ratios of AGNs do not show any correlation with $L_{ir}$. In the upper panel of Figure \ref{fig:agn_excite}, we show [OIII](5008\AA)/H$\beta$ ratio as a function of  $L_{ir}$. In the lower panel of  Figure \ref{fig:agn_excite}, we show  [NII](6585\AA)/H$\alpha$ ratio against $L_{ir}$. In both panels, there are no good correlation between the excitation ratios and $L_{ir}$. The results suggest that the excitation of AGNs are not related to the total power of AGNs.

\subsection{SFR of infrared luminous star-forming galaxies}

 In this subsection, we compare SFR estimated from the optical emission line (H$\alpha$) with $L_{ir}$. Although it has been often suggested that $L_{ir}$ is a good indicator of the  galaxy SFR since the infrared is thermal radiation from dust heated by the HII regions (Condon 1992).  However, the correlation has never been tested with thousands of galaxies. 
 
 We measure optical SFR using the prescription given in Hopkins et al. (2003) and Goto (2005a). In order to correct for stellar absorption, we used H$\delta$ equivalent width (EW) measured in Stoughton et al.(2002), and converted it to the stellar absorption at the H$\alpha$ and H$\beta$ wavelengths using the following empirical relation (Miller \& Owen 2002): H$\beta$ EW$_{absorption}$  is equal to H$\delta$ EW$_{absorption}$ and H$\alpha$ EW$_{absorption}$ is equal to 1.3 + $0.4\times$  H$\delta$ EW$_{absorption}$ (Keel 1983).  Using the stellar absorption corrected H$\alpha$/H$\beta$ flux ratio, we correct for dust extinction using the prescription given in Hopkins et al. (2003). We apply this extinction correction only when both  H$\alpha$ and H$\beta$ lines are detected in emission. Finally we apply the 3" fiber aperture correction by scaling the H$\alpha$ flux by the ratio of the Petrosian $r$-band flux to that measured within the 3'' fiber of the SDSS (Kewley, Jansen, \& Geller 2005). 
 By using this stellar absorption, dust extinction, aperture corrected H$\alpha$ flux to the equation given in Kennicutt (1998), we obtain a SFR for each galaxy. 

\begin{figure}
\begin{center}
\includegraphics[scale=0.4]{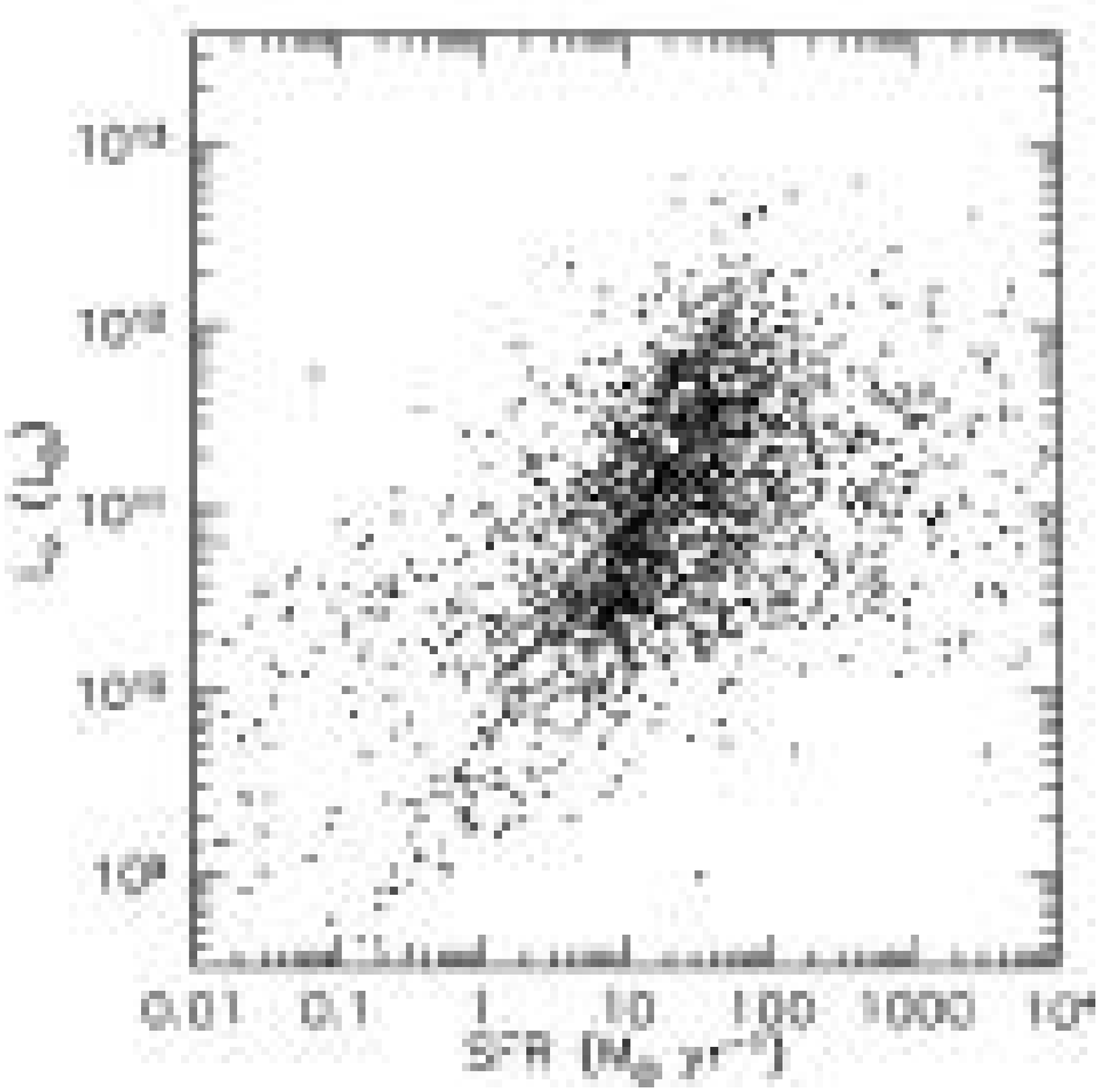}
\end{center}
\caption{The SFR computed from the H$\alpha$ emission line is plotted against the total infrared luminosity ($L_{ir}$). AGNs are excluded from this figure using Fig.\ref{fig:Osterbrook}. 
}\label{fig:sfr_lir} 
\end{figure}

 In Figure \ref{fig:sfr_lir}, we plot the SFR against $L_{ir}$. A good correlation can be found between the SFR and $L_{ir}$, suggesting $L_{ir}$ can be used as a good indicator of a total SFR of the galaxy. The result also supports the idea that in star-forming galaxies,  $L_{ir}$ is mainly from the re-emission of the light of young stars absorbed by the dust (Hopkins et al. 2003; Flores et al. 2004). Note that outliers with unusually large SFR are often nearby galaxies with large aperture and dust corrections.

\begin{figure}
\begin{center}
\includegraphics[scale=0.4]{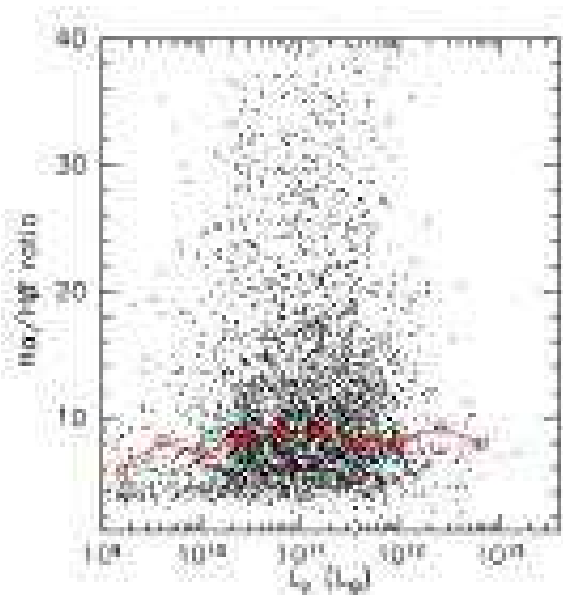}
\end{center}
\caption{
 The self-absorption corrected H$\alpha$/H$\beta$ ratio is plotted against $L_{ir}$.  The solid line connects median values.
}\label{fig:excite}
\end{figure}

 By utilizing the information on H$\alpha$/H$\beta$ flux ratio used in computing the SFR, we can estimate the amount of dust extinction in IRGs. 
In case of CASE B recombination with no dust extinction, the theoretical value of the H$\alpha$/H$\beta$ ratio is 2.87. 
 We plot  H$\alpha$/H$\beta$ ratio against $L_{ir}$ in the lower panel of Figure \ref{fig:excite}. The solid line connects median values.  As can be seen in the figure, there is a slight tendency where the H$\alpha$/H$\beta$ ratio increase with increasing $L_{ir}$. 
 However, it is important to note that the tendency is small compared with the scatter at each $L_{ir}$, suggesting that there is a significant variety in the amount of dust extinction of IRGs.

\subsection{Optical morphology of infrared luminous galaxies}\label{sec:morphology}

 In this subsection, we investigate optical morphology of IRGs. As is found in Figure \ref{fig:Osterbrook}, there are 3438 star-forming IRGs in our sample. In addition to the optically measured redshift, all of them have CCD-based morphology in 5 optical bands ($u,g,r,i,$ and $z$) with a pixel size of 0.396 arcsec/pixel, providing us with an unprecedented opportunity to investigate optical morphology of statistical number of IRGs.

\begin{figure}
\begin{center}
\includegraphics[scale=0.4]{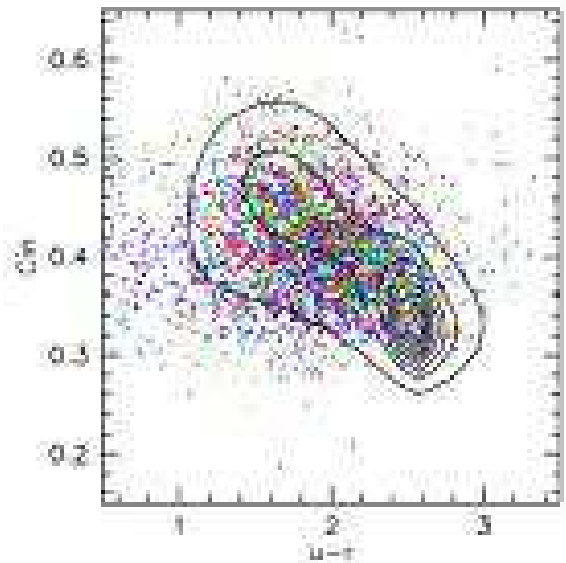}
\end{center}
\caption{The concentration parameter is plotted against restframe $u-r$ colour. The contours represent all galaxies with $r<17.77$ in the SDSS DR3.  }\label{fig:concent}
\end{figure}

In Figure \ref{fig:concent}, a concentration parameter ($Cin$) is plotted against restframe $u-r$ colour. The concentration parameter is computed as a ratio of Petrosian 50\% flux radius to Petrosian 90\% flux radius, and is known to correlate well with traditional eye-based morphological classification (Shimasaku et al. 2001; Strateva et al. 2001). It is also known that restframe $u-r$ colour separates early- and late-type galaxies well (Strateva et al. 2001; Baldry et al. 2004).  The contours denote all non-AGN galaxies in the SDSS DR3 with $r<17.77$ regardless of the IRAS detection. As the contour shows, early-type galaxies have a peak around ($u-r,Cin$)=(2.5,0.35). Late-type galaxies have a peak around ($u-r,Cin$)=(1.8,0.45).  The magenta, blue, green, and red dots are for ULIRGs, VLIRGs, MLIRGs, and LLIRGs, respectively. AGNs are excluded from this plot.

\begin{figure}
\begin{center}
\includegraphics[scale=0.35]{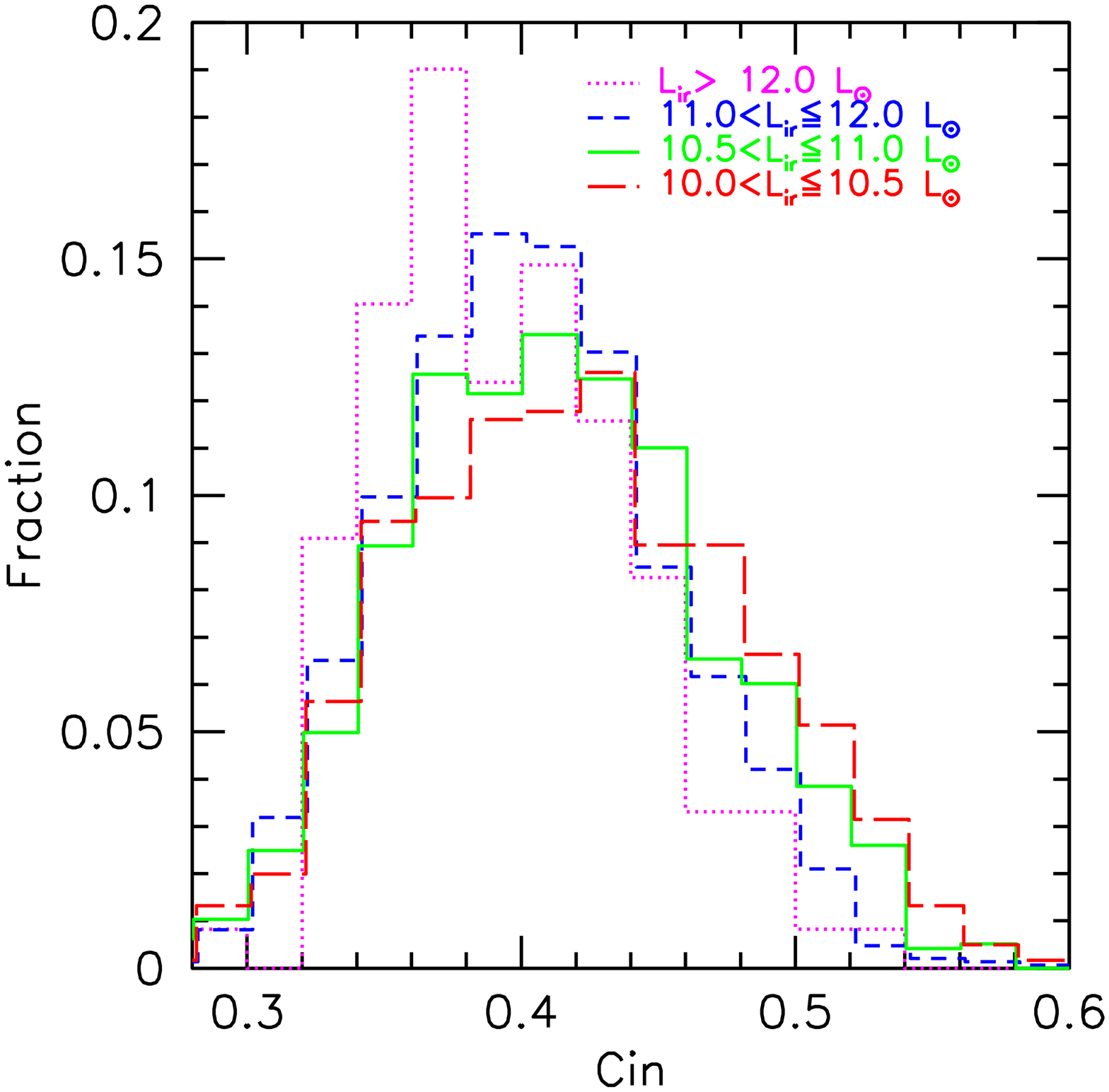}
\includegraphics[scale=0.35]{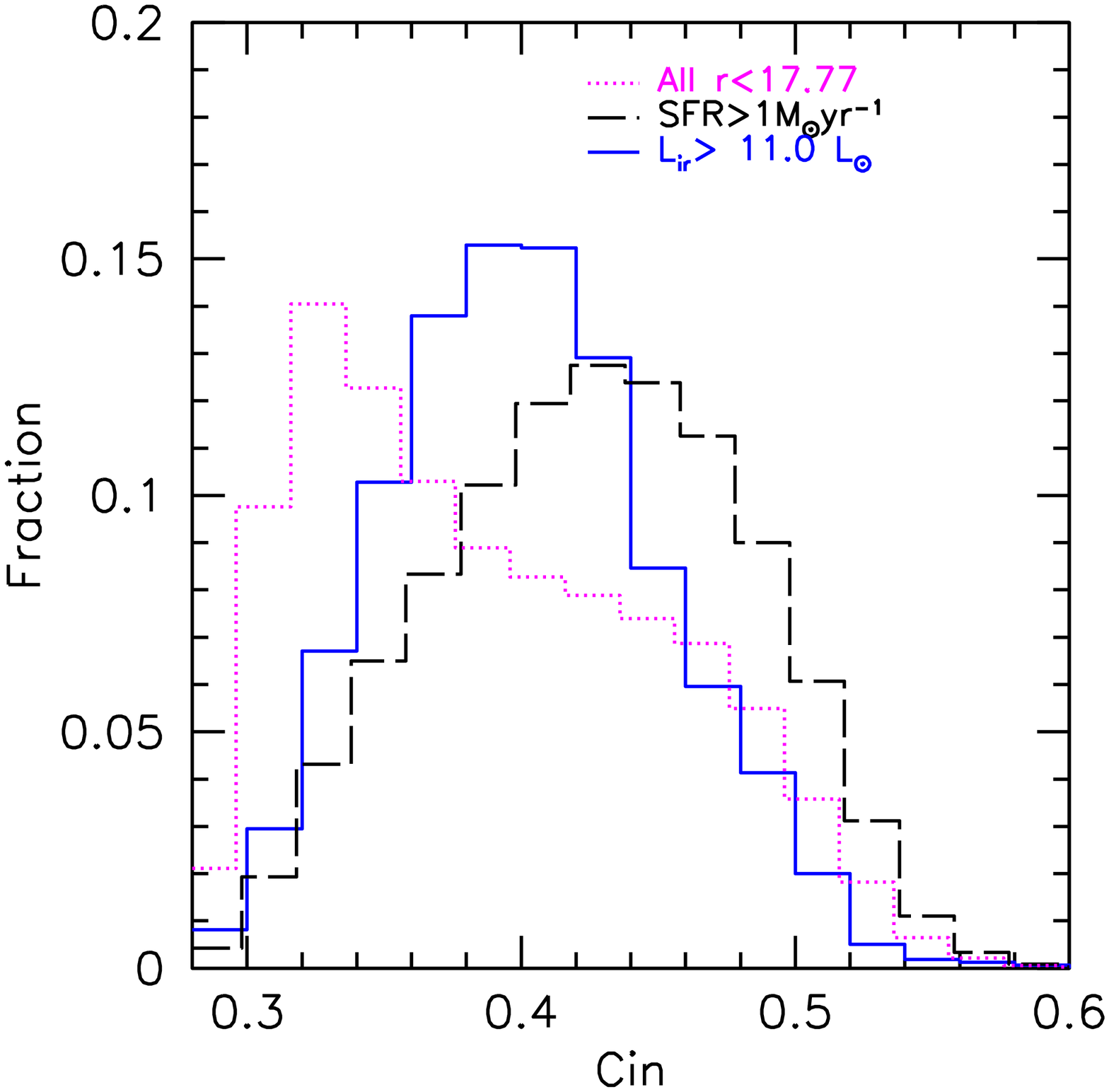}
\includegraphics[scale=0.35]{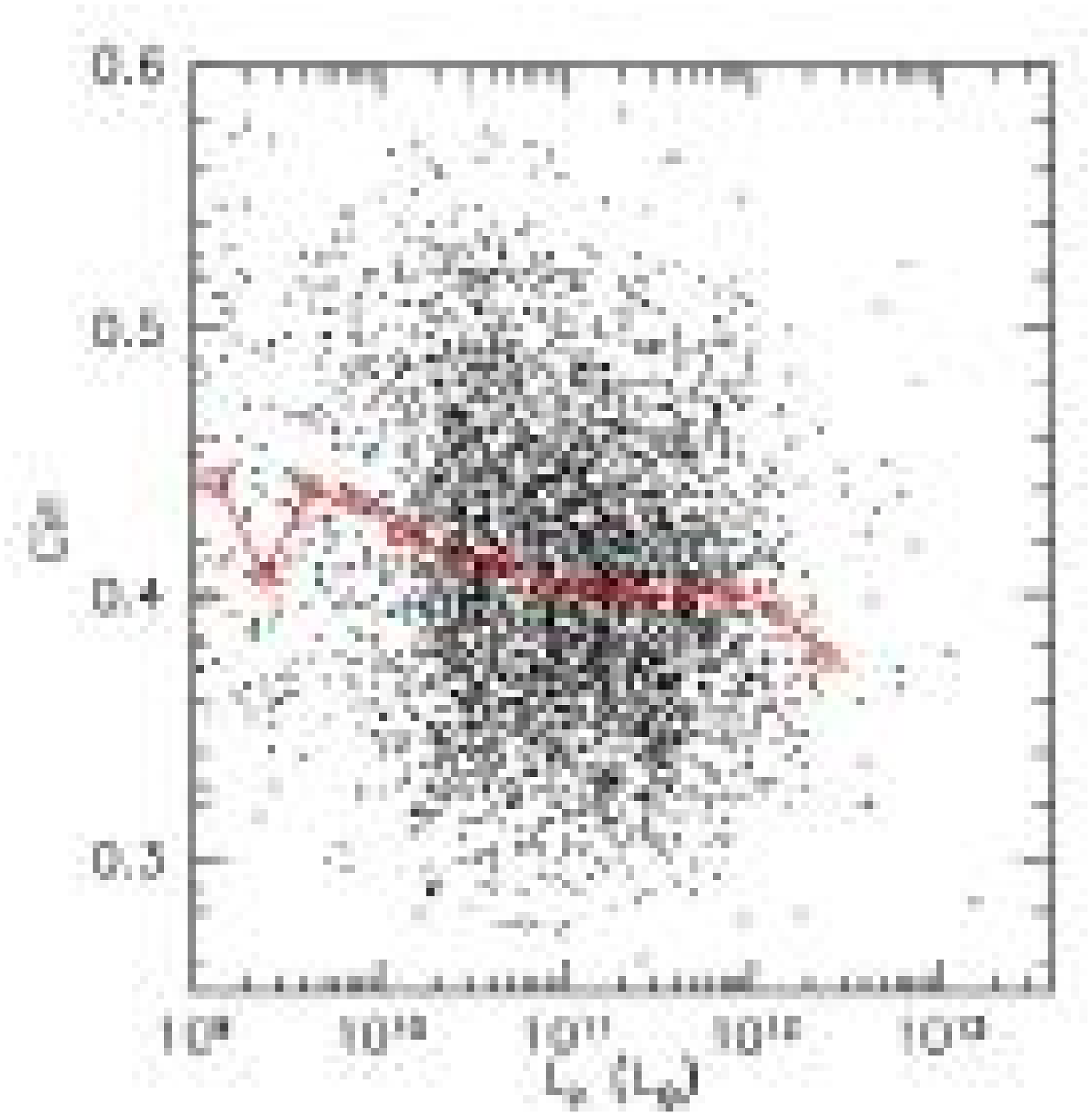}
\end{center}
\caption{Normalized distribution of the concentration parameters.
 In the upper panel, we compare the distribution for ULIRG (dotted line), VLIRGs (short-dashed line), MLIRGs (solid line). and LLIRGs (short-dashed line).
 In the middle panel, the short-dashed line, the long-dashed line and the solid line are for  all galaxies with $r<17.77$, galaxies with SFR$>1M_{\odot}yr^{-1}$, and galaxies with $L_{ir}>10^{11.0}L_{\odot}$. In the bottom panel, $Cin$ is plotted against $L_{ir}$. The solid line connects the median values.}\label{fig:cin_hist}
\end{figure}

 It is interesting to find many IRGs with $Cin\sim 0.35$, indicating significant fractions of IRGs have early-type morphology. In the upper panel of Figure \ref{fig:cin_hist}, we compare the $Cin$ distribution for ULIRGs (dotted line), VLIRGs (short-dashed line), MLIRGs (solid line) and LLIRGs (short-dashed line). It is recognized that more infrared luminous galaxies have more concentrated or earlier-type morphology. This trend is clearer when we plot $Cin$ against $L_{ir}$ in the bottom panel of Figure \ref{fig:cin_hist}. The solid line connects median values. 
It is worrisome if difference in average distance between the samples artificially created this trend. However, our concentration parameter is based on Petrosian parameters, which are  independent of redshift. And thus, the increase of relative size of PSF to physical size of distant object should only make distant objects more extended or less concentrated than nearby objects. In Fig. \ref{fig:cin_hist}, we found distant objects are more concentrated. This trend cannot be caused by the observational bias that more luminous objects are at higher redshift. On the contrary, the non-biased relation may be even steeper than the observed trend in Fig. \ref{fig:cin_hist}.
We further compare the distribution of $Cin$ in the middle panel of Figure \ref{fig:cin_hist}, where we plot the distribution of ULIRGs+VLIRGs with the  solid line and that of all SDSS galaxies with $r<17.77$ with the short-dashed line. In Figure \ref{fig:sfr_lir}, $L_{ir}>10^{11}L_{\odot}$ roughly corresponds to the SFR of 1$M_{\odot}yr^{-1}$. For a comparison, we also plot SDSS galaxies with  the SFR $>1 M_{\odot}yr^{-1}$ with the long-dashed line. 
Very interestingly, although VLIRGs correspond to galaxies with SFR $>1 M_{\odot}yr^{-1}$ (the long-dashed line) in terms of SFR, these two samples have different distributions of $Cin$, in the sense that VLIRGs have more concentrated, or earlier-type morphology. This difference indicates that VLIRGs are not just an infrared-view of  strongly star-forming galaxies, but they might be a different population of galaxies. 

\begin{figure*}
\begin{center}
\includegraphics[scale=0.81]{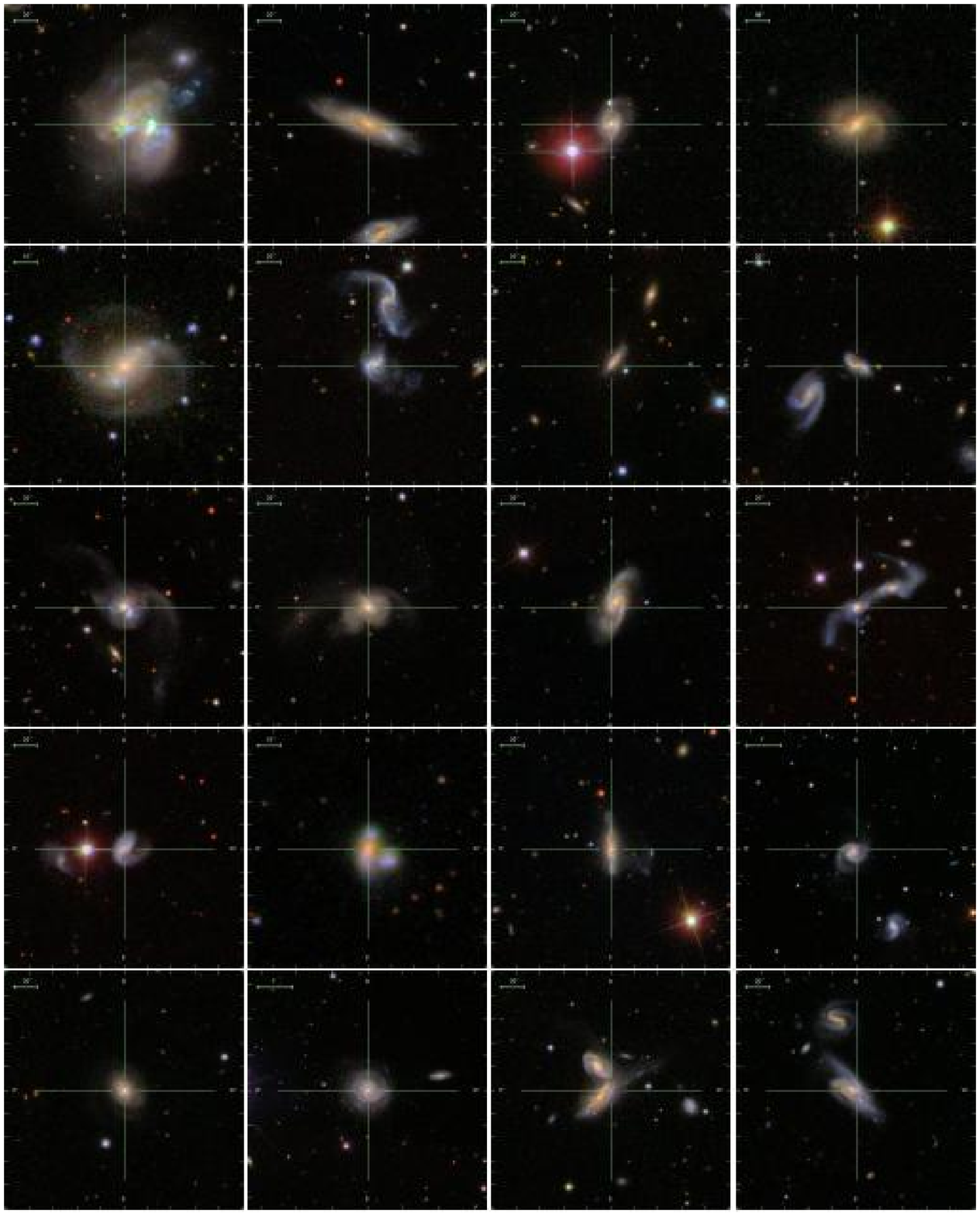}
\end{center}
\caption{Examples of $g,r,i$-composite images of VLIRGs ($10^{11.0}L_{\odot}<L_{ir}$). The images are sorted from low to high redshift. Only 20 lowest redshift galaxies are shown. The corresponding spectra with name and redshift are presented in Fig.\ref{fig:11_spectra}. (The spectrum of the same galaxy can be found in the same column/row panel of Fig.\ref{fig:11_spectra}.)
}\label{fig:11_images}
\end{figure*}

\begin{figure*}
\begin{center}
\includegraphics[scale=0.81]{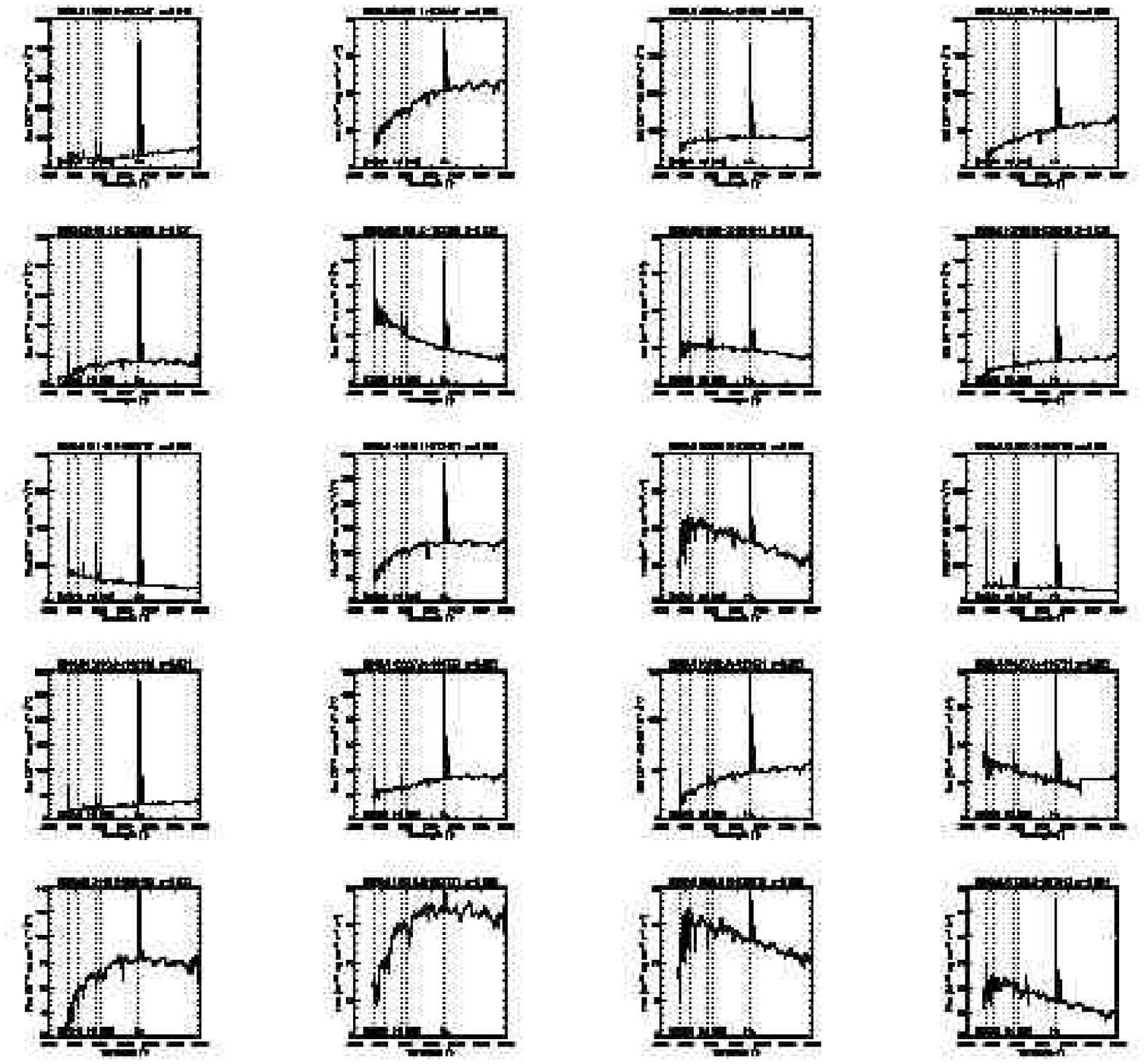}
\end{center}
\caption{Example spectra of 20 lowest redshift VLIRGs. The spectra are sorted from low redshift. Each spectrum is shifted to the restframe wavelength and smoothed using a 20 \AA\ box. The corresponding images are shown in Fig. \ref{fig:11_images}. (The image of the same galaxy can be found in the same column/row panel of Fig.\ref{fig:11_images}).
}\label{fig:11_spectra}
\end{figure*}

\begin{figure*}
\begin{center}
\includegraphics[scale=0.81]{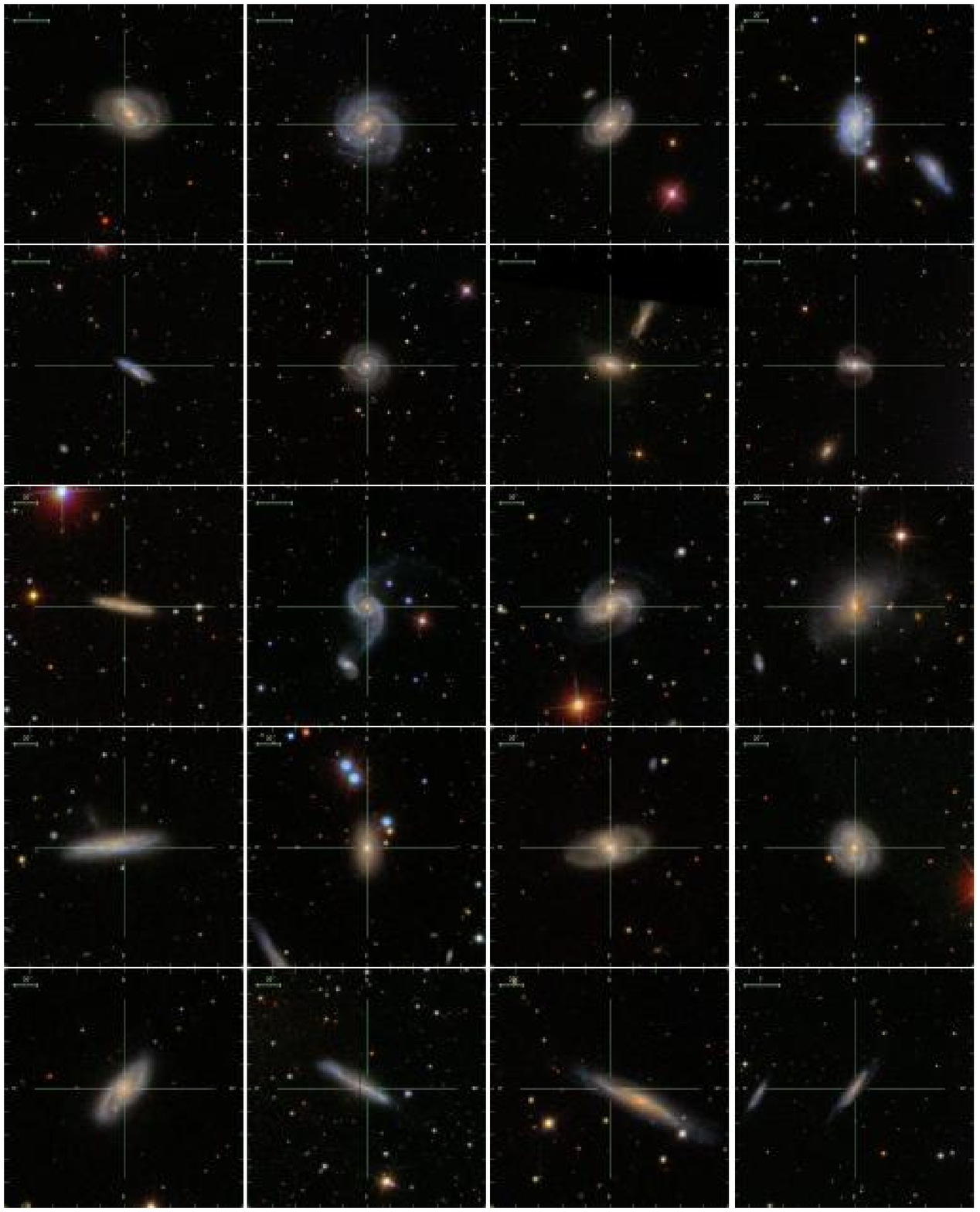}
\end{center}
\caption{Examples of $g,r,i$-composite images of LLIRGs ($10^{10.0}L_{\odot}<L_{ir}\leq 10^{10.5}L_{\odot}$). The images are sorted from low to high redshift. Only 20 lowest redshift galaxies are shown. The corresponding spectra with name and redshift are presented in Fig. \ref{fig:10_spectra}. (The spectrum of the same galaxy can be found in the same column/row panel of Fig.\ref{fig:10_spectra}.)
}\label{fig:10_images}
\end{figure*}

\begin{figure*}
\begin{center}
\includegraphics[scale=0.81]{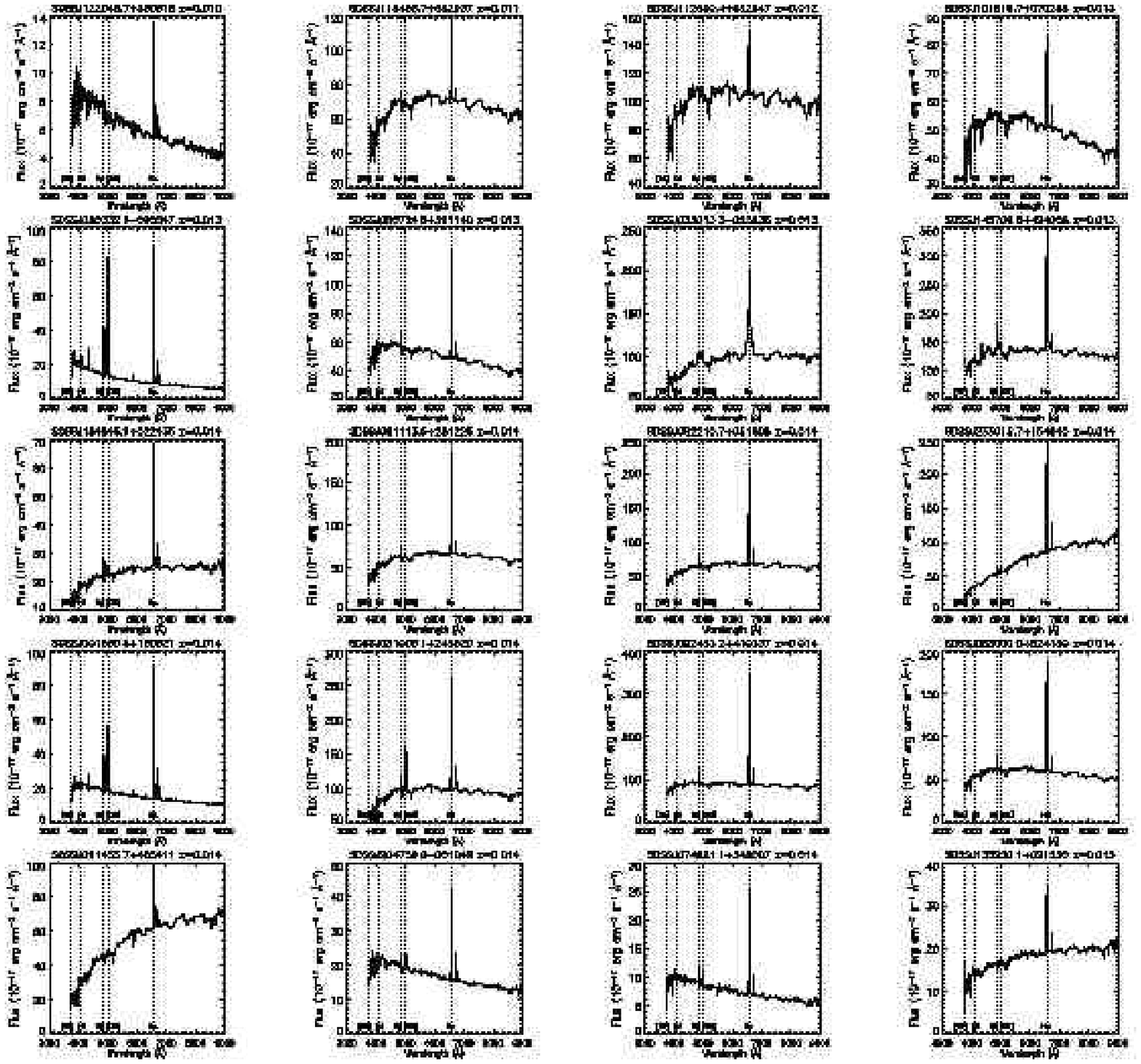}
\end{center}
\caption{Example spectra of 20 lowest redshift LLIRGs. The spectra are sorted from low redshift. Each spectrum is shifted to the restframe wavelength and smoothed using a 20 \AA\ box. The corresponding images are shown in Fig. \ref{fig:10_images}. (The image of the same galaxy can be found in the same column/row panel of Fig.\ref{fig:10_images}).
}\label{fig:10_spectra}
\end{figure*}

This difference manifests itself more when we look at atlas images of IRGs.  In Figure \ref{fig:12_images}, we show example $g,r,i$-composite images of ULIRGs. Nine lowest redshift ULIRGs are selected due to their larger apparent size on the sky. As is previously reported, majority of ULIRGs show dynamically disturbed signatures. The corresponding optical spectra are shown in Figure \ref{fig:12_spectra}.In Figure \ref{fig:11_images}, we show example $g,r,i$-composite images of VLIRGs. 20 lowest redshift VLIRGs are selected. In addition to ULIRGs, most of VLIRGs also show dynamically disturbed morphology. The corresponding optical spectra are shown in Figure \ref{fig:11_spectra}.  Similarly, in Figure \ref{fig:10_images}, we show 20 images of LLIRGs, but LLIRGs are  restricted to $z>0.01$ to match the redshift range of images in Figure \ref{fig:11_images}. The corresponding optical spectra are shown in Figure \ref{fig:10_spectra}. From figs. \ref{fig:12_images},\ref{fig:11_images} and \ref{fig:10_images}, the difference between ULIRGs/VLIRGs and LLIRGs is clear: ULIRGs/VLIRGs have frequent merger/interaction or close companion, whereas LLIRGs often have normal spiral morphology, suggesting that more IR luminous galaxies are powered by the merger/interaction.

\begin{figure*}
\begin{center}
\includegraphics[scale=0.81]{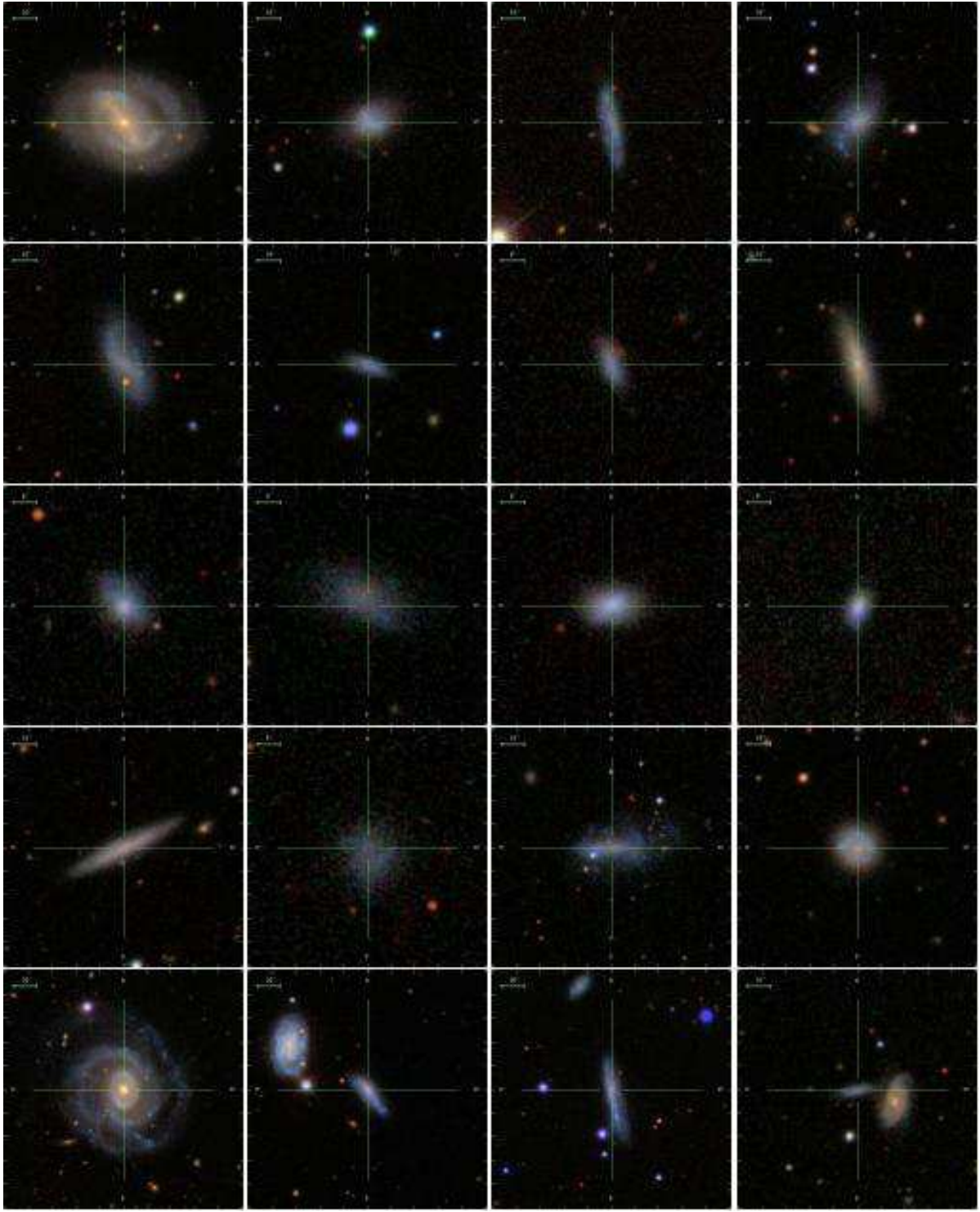}
\end{center}
\caption{Examples of $g,r,i$-composite images of star-forming galaxies ($SFR>1M_{\odot}yr^{-1}$ and $r<17.77$). The images are sorted from low to high redshift. Only 20 lowest redshift galaxies at $z>0.01$ are shown. The corresponding spectra with name and redshift are presented in Fig. \ref{fig:SFG_spectra}. (The spectrum of the same galaxy can be found in the same column/row panel of Fig.\ref{fig:SFG_spectra}.)
}\label{fig:SFG_images}
\end{figure*}

\begin{figure*}
\begin{center}
\includegraphics[scale=0.81]{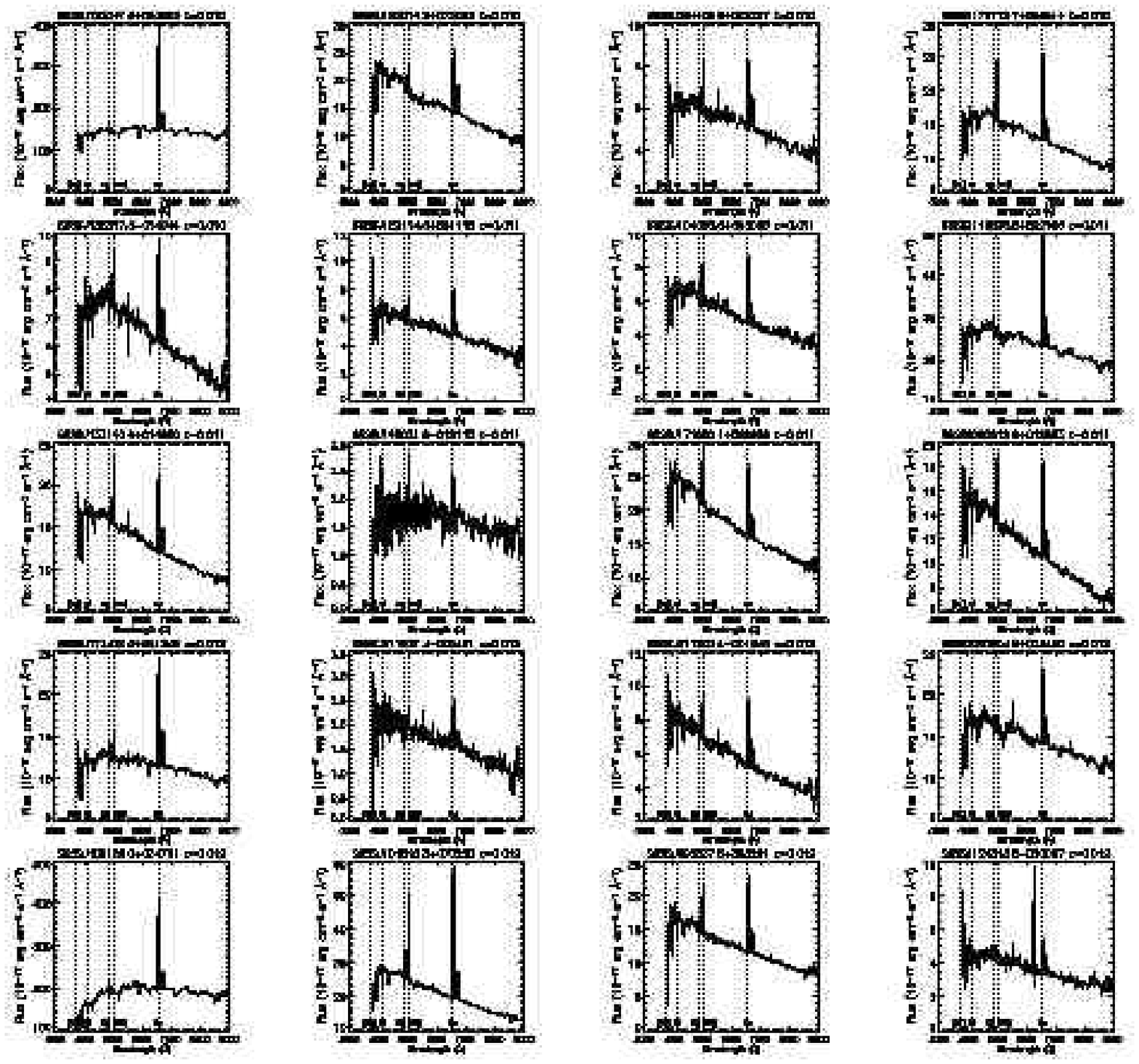}
\end{center}
\caption{Example spectra of 20 lowest redshift star-forming galaxies ($SFR>1M_{\odot}yr^{-1}$ and $r<17.77$) at $z>0.01$. The spectra are sorted from low redshift. Each spectrum is shifted to the restframe wavelength and smoothed using a 20 \AA\ box. The corresponding images are shown in Fig. \ref{fig:SFG_images}. (The image of the same galaxy can be found in the same column/row panel of Fig.\ref{fig:SFG_images}).
}\label{fig:SFG_spectra}
\end{figure*}

In Figure \ref{fig:SFG_images}, we show images of the SDSS galaxies with $SFR>1 M_{\odot}yr^{-1}$ and with $r<17.77$. These galaxies are selected to have similar SFR with VLIRGs shown in Fig.\ref{fig:11_images}. The corresponding optical spectra are shown in Figure \ref{fig:SFG_spectra}. Compared with Fig. \ref{fig:11_images}, although galaxies in both figures have similar SFR of $>1 M_{\odot}yr^{-1}$, there is a marked difference in their morphological appearance: In Fig.\ref{fig:SFG_images}, galaxies are normal spiral or Magellanic-cloud-type irregular galaxies, whereas in Fig.\ref{fig:11_images} majority of galaxies have signs of merger/interaction. The results cautions in estimating total SFR of galaxies using $L_{ir}$, i.e., if we use $L_{ir}$ to estimate SFR of galaxies, we will underestimate the SFR of galaxies that are not experiencing merger/interaction as is shown in Fig.\ref{fig:SFG_images}. Although overall correlation between optically estimated SFR and $L_{ir}$ is good (Fig, \ref{fig:sfr_lir}), careful attention is needed when $L_{ir}$ is used to estimate SFR of individual galaxies.


\subsection{Star Formation History of Infrared Luminous Galaxies}\label{sec:SFH}

\begin{figure}
\begin{center}
\includegraphics[scale=0.4]{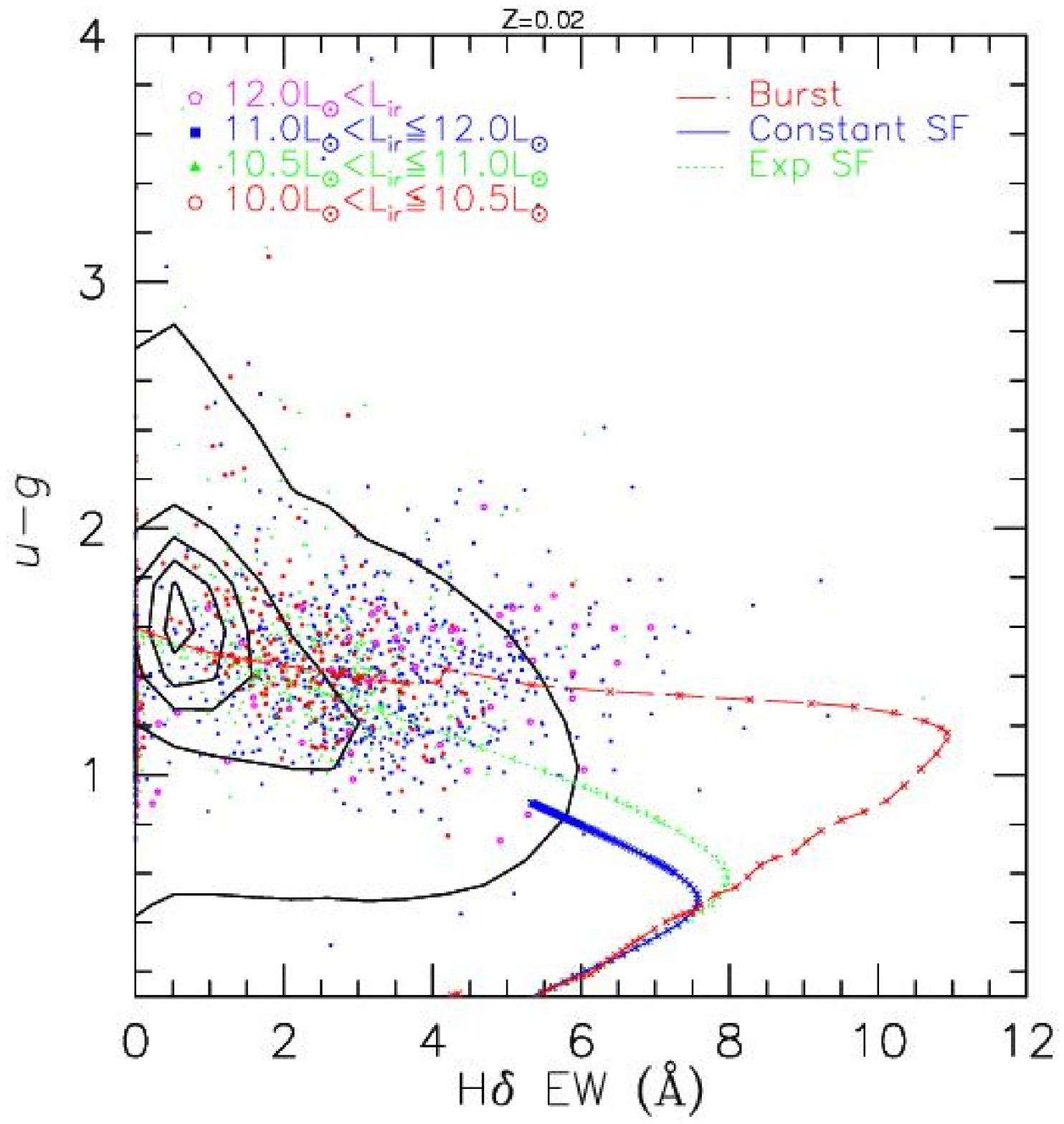}
\end{center}
\caption{Restframe $u-g$ colour is plotted against H$\delta$ EW. The pentagons, squares, triangles, and circles are for ULIRGs, VLIRGs, MLIRGs, and LLIRGs, respectively. The dashed, solid, and dotted lines are for instantaneous burst, constant star formation, and exponentially decaying models ($\tau = 1$Gyr) computed using the GISSEL code (Bruzual \& Charlot 2003). All models assume solar metallicity for simplicity. The contours denote all the SDSS galaxies with $r<17.77$.  }\label{fig:ug_hd}
\end{figure}

In Figure \ref{fig:11_spectra}, we notice that there are significant number of ULIRGs/VLIRGs with strong Balmer absorption lines such as H$\delta  $, which motivates us to investigate star formation history of IRGs. In Figure \ref{fig:ug_hd}, we plot restframe $u-g$ colour against H$\delta$ EW. The $u-g$ colour can be an estimate of the current SFR of galaxies. The H$\delta$ EW is sensitive to the relative amount of A-type stars in the galaxy, and thus, can be an estimate of the past SFR at $\sim$1 Gyr ago. In addition, the broadband $u-g$ colour is much easier to be compared with the population synthesis models than emission lines which are still difficult to be modeled. Therefore, the $u-g$ vs H$\delta$ EW can be used to investigate recent star formation history of IRGs.  In Figure \ref{fig:ug_hd}, the pentagons, squares, triangles, and circles are for ULIRGs, VLIRGs, MLIRGs, and LLIRGs, respectively.  The contours denote all the SDSS galaxies with $r<17.77$. 
 It is recognized that significant fraction of ULIRGs/VLIRGs have slightly redder colour in $u-g$ than the contours especially at the large H$\delta$ EW, indicating the star formation history of VLIRGs may be different from that of normal galaxies. 
 To clarify this further, we over-plot predictions by the population synthesis models (Bruzual \& Charlot 2003). The solid line is for star forming galaxies, where SFR is constant from 0 to 12 Gyr. We use solar metallicity model for simplicity. The dotted line is for exponentially decaying SFR model with $\tau=1$Gyr. This model tracks the center of the contours, and thus, represents a typical star formation history of majority of galaxies. The dashed line denotes instantaneous burst model with the burst duration of 1 Gyr. After the truncation, the galaxy has zero SFR.  This model is known to reproduce the properties of E+A (or post-starburst) galaxies well (Goto et al. 2003c; Goto 2003,2004,2005b). 
 Compared with these models, MLIRGs and LLIRGs are both consistent to be at the end stage of either the exponentially decaying model or the truncation model, suggesting that MLIRGs and LLIRGs are weakening their SFR in a consistent way as normal galaxies. However, ULIRGs/VLIRGs show better agreement with the truncation model than with the exponentially decaying model especially at high H$\delta$ EW end. This may indicate that ULIRGs/VLIRGs are more rapidly reducing their SFR than normal galaxies, at a rate close to that of E+A galaxies. Recently Goto (2005) found that E+A galaxies have more companion galaxies than normal galaxies, and suggested that the origin of E+A galaxies may be dynamical merger/interaction with close companion galaxies. In Section \ref{sec:morphology}, we also found that the origin of ULIRGs/VLIRGs may be dynamical merging of multiple galaxies. If the galaxy-galaxy merging induces the truncation of star-formation, it is naturally explained that both ULIRGs/VLIRGs and E+As are well-reproduced by the truncation model in Fig. \ref{fig:ug_hd}. However, note that perhaps there is no evolutionary link between ULIRGs/VLIRGs and E+A galaxies: ULIRGs/VLIRGs have on-going star formation as observed in FIR or H$\alpha$ emission. On the other hand, E+As have no on-going star formation by definition. Large H$\delta$ EW of ULIRGs/VLIRGs indicate they have already started reducing their SFR. Therefore, it is not likely that ULIRGs/VLIRGs evolve into E+A galaxies which experience a sudden truncation of starburst.

 It is also interesting in terms of evolutionary sequence that we obtained a suggestion that ULIRGs/VLIRGs may be reducing their star formation activity. This can also be intuitively guessed from the presence of strong Balmer absorption lines in ULIRGs/VLIRGs (Fig. \ref{fig:11_spectra}). In the literature, the evolutionary sequence has been often proposed from cool ULIRGs, to warm ULIRGs, to finally QSOs. However, in the sample probed in this work, ULIRGs/VLIRGs seem to be reducing their star formation activity. If so, perhaps  ULIRGs/VLIRGs do not evolve into more vigorous starburst galaxies nor powerful AGNs.

%

%
%


%

%

\section{Discussion}\label{discussion}


 In the literature, there has been much evidence that ULIRGs are caused by the merger/interaction of multiple galaxies: 
 Allen et al. (1985) found 68\% of a sample of 19 optically faint IRAS galaxies were disturbed, interacting, or had close companions.
 Soifer et al. (1986) found that 12-25\% of IRGs with $L_{ir}>10^{11}L_{\odot}$ are peculiar or interacting systems. For 10 ULIRGs, Sanders et al. (1998) concluded that all were interacting or merging. Lawrence et al. (1989) imaged 60 VLIRGs to find 46\% are interacting or merging. The most luminous ULIRGs were universally classified as strong interactions/mergers (Sanders et al. 1987,1988b; Hutchings \& Neff 1987: Vader et al. 1987; Kim 1995; Murphy et al. 1996; Clements et al. 1996). 
 Sanders et al. (1988) suggested that dissipative galaxy merging is important for the formation of ULIRGs. Computationally, numerical simulations by Bekki (1999,2000) successfully reproduced observed $L_{ir}$ of ULIRGs by pair mergers.
 These mergers maybe even multiple mergers of more than two galaxies:
 Taniguchi \& Shioya (1998) theorized that ULIRGs might be a merger of more than two galaxies (also see Xia \& Deng 1997). Bekki et al. (2001) showed that such multiple merging of galaxies can trigger repetitive massive starbursts ($SFR\sim 100 M_{\odot}yr^{-1}$) owing to the strong tidal disturbance and the resultant gaseous dissipation during the merging.
 Indeed, Borne et al. (1999,2000)  observed 99 ULIRGs with HST/WFPC2 in $I$-band to find that 22 of 99 ULIRGs in their sample shows evidence for multiple mergers. Colina et al. (2001) found that the 67\% of the cool ULIRGs observed with HST NICMOS $H$ have multiple nucleus (also see Cui et al. 2001; Bushouse et al. 2002). 
 Our findings of disturbed morphology of ULIRGs in Figure \ref{fig:12_images} are consistent with these results.
 It is not yet clear how much fractions of these merger/interaction is 
by mergers of a pair of galaxies or by multiple galaxies. 
 
According to our work, this merger/interaction origin of IRGs can be applied to lower luminosity VLIRGs. 
  In Section \ref{sec:morphology}, we have found that dynamical merger/interaction between galaxies is more frequent in VLIRGs than in LLIRGs (Figs. \ref{fig:11_images} and \ref{fig:10_images}). 
 It is dangerous to search for the signature of  merger/interaction by eye when one expects to find it. However, in addition to the eye-based scrutiny of the morphological appearance of LIRGs,
 it is revealing that this phenomenon also statistically reveals itself in terms of the concentration parameter ($Cin$), which is computed by a machine, and thus, is much more objective parameter than the eye-based investigation.  We found that more IR luminous galaxies are more centrally concentrated (Fig. \ref{fig:cin_hist}). This is quite a difference from less concentrated morphology of normal star-forming spiral galaxies, suggesting that IR luminous galaxies is a different population of galaxies from normal star-forming spiral galaxies despite of their similarity in the SFR. In Figs. \ref{fig:11_images} and \ref{fig:SFG_images}, there is a distinct difference in morphological appearance between VLIRGs and optical selected star-forming galaxies. Although there is a good correlation between $L_{ir}$ and optical SFR (Fig.\ref{fig:sfr_lir}), the use of  $L_{ir}$ to infer the SFR of galaxies is cautioned. Such estimation may bias high SFR galaxies toward merging/interacting galaxies rather than normal spiral galaxies.  The centrally concentrated appearance of IRGs is consistent with the hypothesis that IRGs, especially ULIRGs/VLIRGs, are created through major merging of multiple galaxies: recent major merger computer simulations suggest that during collisions, the gas readily loses angular momentum due to dynamical friction, decouples from the stars, and inflows rapidly toward the merger nuclei (Barnes \& Hernquist 1992,1996; Mihos \& Hernquist 1994,1996), creating the central starburst, which presumably is observed as high concentration in ULIRGs/VLIRGs (Fig. \ref{fig:cin_hist}). 
 In Fig. \ref{fig:ug_hd}, we found that star formation history of ULIRGs/VLIRGs shows better agreement with the truncation model. This may also support the merger/interaction scenario since it has been suggested that in the case of E+A galaxies  (Zabludoff et al. 1996; Quintero et al. 2004; Blake et al. 2004; Balogh et al. 2004;  Goto 2003,2004,2005b) merger/interaction can induce sudden truncation of star formation.
 In Table \ref{tab:abr}, we have found that IRGs are not very much brighter than the field galaxies in $M_r$. If the origin of VLIRGs is the galaxy-galaxy merger, the merging may not be between two very luminous galaxies ($\gg L^*$), but instead, between less massive sub-$L^*$ galaxies (see also Colina et al. 2001).
  

 The environment of IRGs also shows supportive evidence.
In Section \ref{sec:Mpc}, we found that more infrared luminous galaxies exist in lower density regions. The result is again consistent with the idea that VLIRGs are produced by the merger/interaction of galaxies since such merger/interaction happens more frequently in the field region or in poor groups where relative velocity between galaxies is small. In rich clusters, such merging is less likely to happen since relative velocity is too high ($\sim 1000 km s^{-1}$; Ostriker 1980; Makino \& Hut 1997). 

 In summary, it has been suggested that the origin of ULIRGs/VLIRGs is merger/interaction. In addition to the morphological appearance, our work provided statistical evidence such as concentration parameter, environment of VLIRGs, and star formation history of ULIRGs/VLIRGs to further support the scenario.





\section{Conclusions}\label{conclusion}
We have studied optical properties of unprecedentedly large number of 4248 infrared galaxies by positionally  matching up the SDSS and IRAS data. Our findings can be summarized as follows:
\begin{itemize}
 \item More infrared luminous galaxies tend to exist in lower density regions (Fig. \ref{fig:density}). Especially, VLIRGs have smaller local galaxy density than the field galaxies. Since galaxy-galaxy merging is more frequent in lower density regions where relative velocity of galaxies is smaller, the trend is consistent with the galaxy-galaxy merger origin of VLIRGs.
 \item The fractions of AGNs are higher in more IR luminous galaxies (Fig. \ref{fig:agn_frac}). 
 \item There is a good correlation between $L_{[OIII]}$ and $L_{ir}$, suggesting both parameters can be a good estimator of the total power of AGNs.
 \item There is no significant  correlation between excitation line ratios ([OIII](5008\AA)/H$\beta$ or [NII](6585\AA)/H$\alpha$) and  $L_{ir}$.
 \item There is a good correlation between optically estimated SFR and $L_{ir}$, confirming the previous claim that $L_{ir}$ can be a good indicator of total SFR for star-forming galaxies. However, there is a distinct difference in morphology between optically selected star-forming galaxies and VLIRGs, cautioning that the use of IR estimated SFR may result in a bias toward merging/interacting galaxies having higher SFR than normal spiral galaxies. 
 \item The H$\alpha$/H$\beta$ ratio slightly increases with increasing  $L_{ir}$, suggesting infrared luminous galaxies have more dust extinction. However, the scatter is much larger than the trend, suggesting  that there is a significant variety in the amount of dust extinction of IRGs. 
 \item More infrared luminous galaxies tend to have more concentrated or earlier-type morphology. This is a contrasting result considering that the SFR of VLIRGs corresponds to that of less concentrate spiral galaxies. The result indicates that infrared luminous galaxies are not just an infrared-view of strongly star-forming galaxies, but there is another factor that causes luminous infrared emission. 
 \item Direct comparison of ULIRGs/VLIRGs and LLIRGs images show clear difference: ULIRGs/VLIRGs have more merger/interaction or close companion galaxies, whereas majority of LLIRGs have normal spiral morphology. It is suggested that the merger/interaction with close companions perhaps cause luminous infrared emission in ULIRGs/VLIRGs. 
 \item Comparison with the model SED indicates that ULIRGs/VLIRGs may be in a post-starburst phase, being consistent with the instantaneous burst/truncation model. Considering that the origin of E+A (optically selected post-starburst) galaxies is found to be merger/interaction with close companion galaxies, this may be additional evidence that ULIRGs/VLIRGs may be merging/interacting galaxies. However, ULIRGs/VLIRGs are not likely to be the progenitor of E+A galaxies. The timescale of truncation of star formation is perhaps much shorter for E+A galaxies than ULIRGs/VLIRGs. 

\end{itemize}

%
%
%
%
%


\section*{Acknowledgments}
%
 We thank the anonymous referee for insightful comments, which improved the paper significantly.
    
    Funding for the creation and distribution of the SDSS Archive has been provided by the Alfred P. Sloan Foundation, the Participating Institutions, the National Aeronautics and Space Administration, the National Science Foundation, the U.S. Department of Energy, the Japanese Monbukagakusho, and the Max Planck Society. The SDSS Web site is http://www.sdss.org/.

    The SDSS is managed by the Astrophysical Research Consortium (ARC) for the Participating Institutions. The Participating Institutions are The University of Chicago, Fermilab, the Institute for Advanced Study, the Japan Participation Group, The Johns Hopkins University, Los Alamos National Laboratory, the Max-Planck-Institute for Astronomy (MPIA), the Max-Planck-Institute for Astrophysics (MPA), New Mexico State University, University of Pittsburgh, Princeton University, the United States Naval Observatory, and the University of Washington.




%
%
%




\begin{thebibliography}{DUM}\label{refrence}

%
%
%
%
\bibitem[\protect\citeauthoryear{Abazajian et al.}{2004}]{2004AJ....128..502A} Abazajian K., et al., 2004, AJ, 128, 502 
%

\bibitem[\protect\citeauthoryear{Allen, Roche, \& 
Norris}{1985}]{1985MNRAS.213P..67A} Allen D.~A., Roche P.~F., Norris R.~P., 
1985, MNRAS, 213, 67P 


\bibitem[\protect\citeauthoryear{Arribas et al.}{2004}]{2004AJ....127.2522A} Arribas S., Bushouse H., Lucas R.~A., Colina L., Borne K.~D., 2004, AJ, 127, 2522 



\bibitem[\protect\citeauthoryear{Baldry et al.}{2004}]{2004ApJ...600..681B} 
Baldry I.~K., Glazebrook K., Brinkmann J., Ivezi{\' c} {\v Z}., Lupton 
R.~H., Nichol R.~C., Szalay A.~S., 2004, ApJ, 600, 681 

\bibitem[\protect\citeauthoryear{Balogh et al.}{2004}]{2004ApJ...615L.101B} 
Balogh M.~L., Baldry I.~K., Nichol R., Miller C., Bower R., Glazebrook K., 
2004, ApJ, 615, L101 


\bibitem[\protect\citeauthoryear{Barnes \& 
Hernquist}{1992}]{1992ARA&A..30..705B} Barnes J.~E., Hernquist L., 1992, 
ARA\&A, 30, 705 

\bibitem[\protect\citeauthoryear{Barnes \& 
Hernquist}{1996}]{1996ApJ...471..115B} Barnes J.~E., Hernquist L., 1996, 
ApJ, 471, 115 



\bibitem[\protect\citeauthoryear{Bekki, Shioya, \& 
Tanaka}{1999}]{1999ApJ...520L..99B} Bekki K., Shioya Y., Tanaka I., 1999,
ApJ, 520, L99


\bibitem[\protect\citeauthoryear{Bekki}{2001}]{2001ApJ...546..189B} Bekki 
K., 2001, ApJ, 546, 189

\bibitem[\protect\citeauthoryear{Bekki \&
Shioya}{2001}]{2001ApJS..134..241B} Bekki K., Shioya Y., 2001, ApJS, 134,
241
%
%
%
%
%
\bibitem[\protect\citeauthoryear{Bekki, Shioya, \& 
Couch}{2001}]{2001ApJ...547L..17B} Bekki K., Shioya Y., Couch W.~J., 2001, 
ApJ, 547, L17 
%
%
%
%
\bibitem[\protect\citeauthoryear{Bennett et al.}{2003}]{2003ApJS..148....1B} Bennett C.~L., et al., 2003, ApJS, 148, 1 
%

\bibitem[\protect\citeauthoryear{Blake et al.}{2004}]{2004MNRAS.355..713B} 
Blake C., et al., 2004, MNRAS, 355, 713 


\bibitem[]{2003AJ....125.2348B} Blanton 
M.~R., Brinkmann J., Csabai I., et al., 2003a, AJ,  125, 2348 
%

\bibitem[\protect\citeauthoryear{Borne et al.}{1999}]{1999Ap&SS.266..137B} 
Borne K.~D., et al., 1999, Ap\&SS, 266, 137 

\bibitem[\protect\citeauthoryear{Borne et al.}{2000}]{2000ApJ...529L..77B} 
Borne K.~D., Bushouse H., Lucas R.~A., Colina L., 2000, ApJ, 529, L77 



\bibitem[\protect\citeauthoryear{Bruzual \& 
Charlot}{2003}]{2003MNRAS.344.1000B} Bruzual G., Charlot S., 2003, MNRAS, 
344, 1000 

%
%
%
%

\bibitem[\protect\citeauthoryear{Bushouse et 
al.}{2002}]{2002ApJS..138....1B} Bushouse H.~A., et al., 2002, ApJS, 138, 1 

%
%
%
%
%
%
%


\bibitem[\protect\citeauthoryear{Clements et 
al.}{1996}]{1996MNRAS.279..477C} Clements D.~L., Sutherland W.~J., McMahon 
R.~G., Saunders W., 1996, MNRAS, 279, 477 

\bibitem[\protect\citeauthoryear{Cohen et al.}{1987}]{1987AJ.....93.1199C} 
Cohen M., Schwartz D.~E., Chokshi A., Walker R.~G., 1987, AJ, 93, 1199 

\bibitem[\protect\citeauthoryear{Colina et al.}{2001}]{2001ApJ...563..546C} 
Colina L., et al., 2001, ApJ, 563, 546 

 \bibitem[]{1992ARA&A..30..575C} Condon J.~J., 
1992, ARA\&A,  30, 575 
%
%
%
\bibitem[\protect\citeauthoryear{Cui et al.}{2001}]{2001AJ....122...63C} 
Cui J., Xia X.-Y., Deng Z.-G., Mao S., Zou Z.-L., 2001, AJ, 122, 63 

\bibitem[\protect\citeauthoryear{Doyon, Joseph, \& 
Wright}{1994}]{1994ApJ...421..101D} Doyon R., Joseph R.~D., Wright G.~S., 
1994, ApJ, 421, 101 

%
%
%
%
%
%
%
%
%

\bibitem[\protect\citeauthoryear{Flores et al.}{2004}]{2004A&A...415..885F} 
Flores H., Hammer F., Elbaz D., Cesarsky C.~J., Liang Y.~C., Fadda D., 
Gruel N., 2004, A\&A, 415, 885 

%
%
%
%
%

\bibitem[\protect\citeauthoryear{Genzel et al.}{2001}]{2001ApJ...563..527G} 
Genzel R., Tacconi L.~J., Rigopoulou D., Lutz D., Tecza M., 2001, ApJ, 563, 
527 

%
%
\bibitem[\protect\citeauthoryear{Goto et al.}{2002}]{2002PASJ...54..515G} 
Goto T., et al., 2002a, PASJ, 54, 515
%
\bibitem[\protect\citeauthoryear{Goto et al.}{2002}]{2002AJ....123.1807G} 
Goto T., et al., 2002b, AJ, 123, 1807
%
%
%
%
\bibitem[\protect\citeauthoryear{Goto}{2003}]{2003PhDT.........2G} Goto T.,
2003, PhD Thesis, The University of Tokyo, astro-ph/0310196

%
%

%
%
%
%
%
%
\bibitem[\protect\citeauthoryear{Goto et al.}{2003}]{2003MNRAS.346..601G}
Goto T., Yamauchi C., Fujita Y., Okamura S., Sekiguchi M., Smail I.,
Bernardi M., Gomez P.~L., 2003a, MNRAS, 346, 601 
%
\bibitem[\protect\citeauthoryear{Goto et al.}{2003}]{2003PASJ...55..739G} 
Goto T., et al., 2003b, PASJ, 55, 739  

\bibitem[\protect\citeauthoryear{Goto et al.}{2003}]{2003PASJ...55..771G} Goto T., et al., 2003c, PASJ, 55, 771 

%
%


\bibitem[\protect\citeauthoryear{Goto}{2004}]{2004A&A...427..125G} Goto T., 
2004, A\&A, 427, 125 

%



\bibitem[\protect\citeauthoryear{Goto}{2005}]{2005MNRAS.356L...6G} Goto T., 
2005a, MNRAS, 356, L6 


\bibitem[\protect\citeauthoryear{Goto}{2005}]{2005MNRAS.357..937G} Goto T., 
2005b, MNRAS, 357, 937 
%
%
%
%
%
%
%
%
%
%


\bibitem[\protect\citeauthoryear{Hopkins et 
al.}{2003}]{2003ApJ...599..971H} Hopkins A.~M., et al., 2003, ApJ, 599, 971 

\bibitem[\protect\citeauthoryear{Hutchings \& 
Neff}{1987}]{1987AJ.....93...14H} Hutchings J.~B., Neff S.~G., 1987, AJ, 
93, 14 




\bibitem[\protect\citeauthoryear{Keel}{1983}]{1983ApJ...269..466K} Keel 
W.~C., 1983, ApJ, 269, 466 

\bibitem[]{1998ARA&A..36..189K} Kennicutt R.~C., 1998, ARA\&A,  36, 189 


\bibitem[]{2001K} Kewley L.~J., Dopita M.~A., Sutherland R.~S., Heisler C.~A., Trevena J., 2001, ApJ,  556, 121 



\bibitem[]{2005K}Kewley L.~J., Jansen R. A.,  Geller M. J., 2005 PASP in press, astro-ph/0501229


\bibitem[\protect\citeauthoryear{Kim, Veilleux, \& Sanders}{2002}]{2002ApJS..143..277K} Kim D.-C., Veilleux S., Sanders D.~B., 2002, ApJS, 143, 277 

\bibitem[\protect\citeauthoryear{Kim, Veilleux, \& Sanders}{1998}]{1998ApJ...508..627K} Kim D.-C., Veilleux S., Sanders D.~B., 1998, ApJ, 508, 627 

\bibitem[\protect\citeauthoryear{Kim \& Sanders}{1998}]{1998ApJS..119...41K} Kim D.-C., Sanders D.~B., 1998, ApJS, 119, 41 

\bibitem[\protect\citeauthoryear{Kim et al.}{1995}]{1995ApJS...98..129K} 
Kim D.-C., Sanders D.~B., Veilleux S., Mazzarella J.~M., Soifer B.~T., 
1995, ApJS, 98, 129 


\bibitem[\protect\citeauthoryear{Lawrence et 
al.}{1989}]{1989MNRAS.240..329L} Lawrence A., Rowan-Robinson M., Leech K., 
Jones D.~H.~P., Wall J.~V., 1989, MNRAS, 240, 329 

\bibitem[\protect\citeauthoryear{Leech et al.}{1989}]{1989MNRAS.240..349L} 
Leech K.~J., Penston M.~V., Terlevich R., Lawrence A., Rowan-Robinson M., 
Crawford J., 1989, MNRAS, 240, 349 




\bibitem[\protect\citeauthoryear{Lutz et al.}{1998}]{1998ApJ...505L.103L} 
Lutz D., Spoon H.~W.~W., Rigopoulou D., Moorwood A.~F.~M., Genzel R., 1998, 
ApJ, 505, L103 


\bibitem[]{1997ApJ...481...83M} Makino J., Hut P., 1997, ApJ,  481, 83 
%
%
%

\bibitem[\protect\citeauthoryear{Mihos \& 
Hernquist}{1996}]{1996ApJ...464..641M} Mihos J.~C., Hernquist L., 1996, 
ApJ, 464, 641 

\bibitem[\protect\citeauthoryear{Mihos \& 
Hernquist}{1994}]{1994ApJ...431L...9M} Mihos J.~C., Hernquist L., 1994, 
ApJ, 431, L9 

%
\bibitem[]{2002AJ....124.2453M} Miller N.~A., Owen F.~N., 2002, AJ,  124, 2453 

 \bibitem{}Moshir, M., et al. 1992, Explanatory Supplement to the IRAS Faint Source Survey, Version 2, JPL D-10015 8/92 (Pasadena: JPL)


\bibitem[\protect\citeauthoryear{Murphy et al.}{1996}]{1996AJ....111.1025M} 
Murphy T.~W., Armus L., Matthews K., Soifer B.~T., Mazzarella J.~M., Shupe 
D.~L., Strauss M.~A., Neugebauer G., 1996, AJ, 111, 1025 

\bibitem[\protect\citeauthoryear{Nagar et al.}{2003}]{2003A&A...409..115N} 
Nagar N.~M., Wilson A.~S., Falcke H., Veilleux S., Maiolino R., 2003, A\&A, 
409, 115 



\bibitem[\protect\citeauthoryear{Neugebauer et 
al.}{1984}]{1984ApJ...278L...1N} Neugebauer G., et al., 1984, ApJ, 278, L1 

\bibitem[]{1980ComAp...8..177O} Ostriker 
J.~P., 1980, ComAp,  8, 177 
%
%
\bibitem[\protect\citeauthoryear{Quintero et 
al.}{2004}]{2004ApJ...602..190Q} Quintero A.~D., et al., 2004, ApJ, 602, 
190 

\bibitem[Sanders et al.(1988)]{1988ApJ...325...74S} Sanders, D.~B., Soifer,
B.~T., Elias, J.~H., Madore, B.~F., Matthews, K., Neugebauer, G., \&
Scoville, N.~Z.\ 1988, \apj, 325, 74

\bibitem[\protect\citeauthoryear{Sanders \& 
Mirabel}{1996}]{1996ARA&A..34..749S} Sanders D.~B., Mirabel I.~F., 1996, 
ARA\&A, 34, 749 


\bibitem[\protect\citeauthoryear{Sanders}{2003}]{2003JKAS...36..149S} 
Sanders D.~B., 2003, JKAS, 36, 149

\bibitem[\protect\citeauthoryear{Sarazin}{1976}]{1976ApJ...208..323S} 
Sarazin C.~L., 1976, ApJ, 208, 323 


\bibitem[\protect\citeauthoryear{Scoville et 
al.}{2000}]{2000AJ....119..991S} Scoville N.~Z., et al., 2000, AJ, 119, 991 


\bibitem[\protect\citeauthoryear{Stanford \& 
Bushouse}{1991}]{1991ApJ...371...92S} Stanford S.~A., Bushouse H.~A., 1991, 
ApJ, 371, 92 

%
%
\bibitem[]{2001AJ....122.1238S} Shimasaku K., Fukugita M., Doi M., et al., 2001, AJ,  122, 1238 


\bibitem[\protect\citeauthoryear{Soifer et al.}{1984}]{1984ApJ...278L..71S} 
Soifer B.~T., et al., 1984, ApJ, 278, L71 

\bibitem[\protect\citeauthoryear{Soifer et al.}{1987}]{1987ApJ...320..238S} 
Soifer B.~T., Sanders D.~B., Madore B.~F., Neugebauer G., Danielson G.~E., 
Elias J.~H., Lonsdale C.~J., Rice W.~L., 1987, ApJ, 320, 238 


\bibitem[]{2002AJ....123..485S} Stoughton 
C., Lupton R.~H., Bernardi M., et al., 2002, AJ,  123, 485 
%
\bibitem[\protect\citeauthoryear{Strateva et al.}{2001}]{2001AJ....122.1861S} Strateva I., et al., 2001, AJ, 122, 1861 
%
%
%
%


\bibitem[\protect\citeauthoryear{Taniguchi \& 
Shioya}{1998}]{1998ApJ...501L.167T} Taniguchi Y., Shioya Y., 1998, ApJ, 
501, L167 

%
%

\bibitem[\protect\citeauthoryear{Vader et al.}{1987}]{1987AJ.....94..847V} 
Vader J.~P., Da Costa G.~S., Heisler C.~A., Frogel J.~A., Simon M., 1987, 
AJ, 94, 847 


\bibitem[\protect\citeauthoryear{Veilleux, Kim, \& 
Sanders}{2002}]{2002ApJS..143..315V} Veilleux S., Kim D.-C., Sanders D.~B., 
2002, ApJS, 143, 315 

\bibitem[\protect\citeauthoryear{Veilleux, Kim, \& 
Sanders}{1999}]{1999ApJ...522..113V} Veilleux S., Kim D.-C., Sanders D.~B., 
1999, ApJ, 522, 113 

\bibitem[\protect\citeauthoryear{Veilleux}{1999}]{1999Ap&SS.266...67V} 
Veilleux S., 1999, Ap\&SS, 266, 67 



\bibitem[\protect\citeauthoryear{Wright et al.}{1990}]{1990Natur.344..417W} Wright G.~S., James P.~A., Joseph R.~D., McLean I.~S., 1990, Natur, 344, 417 

\bibitem[\protect\citeauthoryear{Wu et al.}{1998}]{1998A&AS..132..181W} Wu
H., Zou Z.~L., Xia X.~Y., Deng Z.~G., 1998, A\&AS, 132, 181

\bibitem[\protect\citeauthoryear{Wu et al.}{1998}]{1998A&AS..127..521W} Wu
H., Zou Z.~L., Xia X.~Y., Deng Z.~G., 1998, A\&AS, 127, 521


%
%
\bibitem[Zabludoff et al.(1996)]{1996ApJ...466..104Z} Zabludoff, A.~I.,
Zaritsky, D., Lin, H., Tucker, D., Hashimoto, Y., Shectman, S.~A., Oemler,
A., \& Kirshner, R.~P.\ 1996, \apj, 466, 104
\end{thebibliography}
\end{document}